\documentclass[journal]{IEEEtran}

\ifCLASSINFOpdf
\usepackage[pdftex]{graphicx}
\graphicspath{{./Images/}}
\usepackage{epstopdf}
\else
\fi
\usepackage{cite}
\usepackage{amsmath,amssymb,amsfonts}
\usepackage{graphicx}
\usepackage{textcomp}
\usepackage{color}
\usepackage{siunitx}
\usepackage{enumitem}
\DeclareRobustCommand{\rev}[1]{%
	\ifmmode
	\begingroup
	\color{blue}%
	#1%
	\endgroup
	\else
	\textcolor{blue}{#1}%
	\fi
}

\DeclareSIUnit \voltampere { VA } %apparent power 
\DeclareSIUnit \var { VAr } %volt-ampere reactive - idle power .
\DeclareMathAlphabet\mathbfcal{OMS}{cmsy}{b}{n}
\usepackage[acronym]{glossaries}
\usepackage{nomencl}
\makenomenclature
\usepackage{etoolbox}
\renewcommand\nomgroup[1]{%
	\item[\bfseries	
		\ifstrequal{#1}{A}{Acronyms}{%
			\ifstrequal{#1}{S}{Sets}{%
			\ifstrequal{#1}{D}{Decision Variables}{%
				\ifstrequal{#1}{P}{Parameters}{}}{%
					\ifstrequal{#1}{I}{Initial Conditions}{}}}}%
	]}

\begin{document}
	
	\title{Modelling and Characterisation of Flexibility\\from Distributed Energy Resources}
	
	%\vspace{-0.5em}
	\author{Shariq~Riaz,~\IEEEmembership{Senior Member,~IEEE,}
		and~Pierluigi~Mancarella,~\IEEEmembership{Senior Member,~IEEE}% <-this % stops a space
		%\vspace{-1.5em}
		\thanks{The authors gratefully acknowledge the partial support for this research received from the Victorian Government’s \emph{veski} program and John Theunissen from Ausnet Services for the insightful discussions on the topic.}
		
		\thanks{Shariq~Riaz and Pierluigi~Mancarella are with The University of Melbourne, Parkville, 3010, Victoria, Australia. E-mails:\{shariq.riaz, pierluigi.mancarella\}@unimelb.edu.au. Pierluigi~Mancarella is also with the Department of Electrical and Electronic Engineering,
		The University of Manchester, Manchester M13 9PL, U.K. E-mail:
		p.mancarella@manchester.ac.uk.}% <-this % stops a space
		%\thanks{J. Doe and J. Doe are with Anonymous University.}% <-this % stops a space

	}
	\maketitle
	
	% As a general rule, do not put math, special symbols or citations
	% in the abstract or keywords.
	
	\begin{abstract}
		Harnessing flexibility from distributed energy resources (DER) to participate in various markets while accounting for relevant technical and commercial constraints is essential for the development of low-carbon grids. However, there is no clear definition or even description of the salient features of aggregated DER flexibility, including its steady-state and dynamic features and how these are impacted by network constraints and market requirements. This paper proposes a comprehensive DER flexibility modelling and characterisation framework that is based on the concept of \emph{nodal operating envelope} (NOE). In particular, capacity, ramp, duration and cost are identified as key flexibility metrics and associated with different but consistent NOEs describing \emph{capability}, \emph{feasibility}, \emph{ramp}, \emph{duration}, \emph{economic}, \emph{technical} and \emph{commercial flexibility} features. These NOEs, which conceptually arise from a Venn diagram, can be built via optimal power flow (OPF) analysis, visualised in the active-reactive power space, and used by different stakeholders. Results from a canonical test system and a real distribution system illustrate the value and applicability of the proposed framework to model and characterise provision of flexibility and market services from DER for different use cases.   
		
	\end{abstract}
	\nomenclature[Avpp]{VPP}{Virtual power plant}
	\nomenclature[Afor]{FOR}{Feasible operating region}
	\nomenclature[Ader1]{DER}{Distributed energy resources}
	\nomenclature[Atso]{TSO}{Transmission system operator}
	\nomenclature[Adso]{DSO}{Distribution system operator}
	\nomenclature[Ader2]{DERA}{DER aggregations}
	\nomenclature[Apq]{P-Q}{Active-reactive power}
	\nomenclature[Anoe]{NOE}{Nodal operating envelope}
	\nomenclature[Aopf]{OPF}{Optimal power flow}
	\nomenclature[Abess]{BESS}{Battery energy storage system}
	\nomenclature[Arfe]{RFE}{Ramp flexibility envelope}
	\nomenclature[Adfe]{DFE}{Duration flexibility envelope}
	\nomenclature[Aefe]{EFE}{Economic flexibility envelope}
	\nomenclature[Atfe]{TFE}{Technical flexibility envelope}
	\nomenclature[Acfe]{CFE}{Commercial flexibility envelope}
	\nomenclature[Admo]{DMO}{Distribution market operator}
	\nomenclature[Aoltc]{OLTC}{On load tap changer}
	\nomenclature[Afcas]{FCAS}{Frequency controlled ancillary services}
	\nomenclature[Asrmc]{SRMC}{Short run marginal cost}
	\nomenclature[Anem]{NEM}{Australian National Electricity Market}
	
	\nomenclature[Sc]{$\mathcal{C}$}{Capability NOE of DERA}
	\nomenclature[Sr]{$\mathcal{R}$}{Set of resources $r$}
	\nomenclature[SSr]{$\mathcal{S}_r$}{Capability set of resource $r$}
	\nomenclature[Dsra]{$\boldsymbol{s}_r$}{dispatch point of resource $r$; $\boldsymbol{s}_r \in \mathcal{S}_r$; $\boldsymbol{s}_r=[p_r,q_r]^T$}
	%\nomenclature[Dsdr]{$\Delta\boldsymbol{s}_r$}{power deviation of resource $r$; $\Delta \boldsymbol{s}_r=[\Delta p_r,\Delta q_r]^T$}
	\nomenclature[Psla]{$\boldsymbol{s}^\lambda$}{DERA dispatch point; $\boldsymbol{s}^\lambda=[p^\lambda,q^\lambda ]^T$}
	\nomenclature[Sf01]{$\mathcal{F}$}{Feasibility NOE of DERA}
	\nomenclature[Psd]{ $\boldsymbol{s}_d$}{Power requirement of demand $d$; $\boldsymbol{s}_d=[p_d,q_d ]^T$}
	\nomenclature[Sd]{$\mathcal{D}$}{Set of demand $d$}
	\nomenclature[Sl]{$\mathcal{L}$}{Set of network elements $l$}
	\nomenclature[Dsl]{ $\boldsymbol{s}_l$}{Power flow in network element $l$; $\boldsymbol{s}_l=[p_l,q_l ]^T$}
	\nomenclature[Dsl]{ $\boldsymbol{s}_l^{\text{loss}}$}{Power loss in network element $l$}
	\nomenclature[Dvi]{$\boldsymbol{v}_i$}{Voltage vector of bus $i$; $\boldsymbol{v}_i= v_i\angle\theta_i$}
	\nomenclature[P01]{$\underline{\bullet}$ / $\overline{\bullet}$}{Minimum/maximum limit of variable $\bullet$}
	\nomenclature[Sc]{$\mathcal{C}$}{Capability NOE of DERA}
	\nomenclature[Sr]{$\mathcal{R}$}{Set of resources $r$}
	\nomenclature[SSr]{$\mathcal{S}_r$}{Capability set of resource $r$}
	\nomenclature[Dsr]{$\boldsymbol{s}_r$}{dispatch point of resource $r$; $\boldsymbol{s}_r \in \mathcal{S}_r$; $\boldsymbol{s}_r=[p_r,q_r]^T$}
	\nomenclature[Dsrd]{$\Delta\boldsymbol{s}_r$}{power deviation of resource $r$; $\Delta \boldsymbol{s}_r=[\Delta p_r,\Delta q_r]^T$}
	\nomenclature[Psgl]{$\boldsymbol{s}^\lambda$}{DERA dispatch point; $\boldsymbol{s}^\lambda=[p^\lambda,q^\lambda ]^T$}
	\nomenclature[P1t]{$\tau$}{Response time}
	\nomenclature[P1p]{$\psi$}{call length}
	\nomenclature[Prr]{$r_r^{\text{+/-}}$}{ramp up/down capability of resourse $r$}
	\nomenclature[Der]{$e_r$}{energy content of  resource $r$}
	\nomenclature[Sfr]{$\mathcal{F}^\text{ramp}$}{Ramp flexiblity NOE}
	\nomenclature[Sfd]{$\mathcal{F}^\text{duration}$}{Duration flexiblity NOE}
	\nomenclature[Sfe]{$\mathcal{F}^\text{economic}$}{Economic flexiblity NOE}
	\nomenclature[Sft]{$\mathcal{F}^\text{technical}$}{Technical flexibility NOE}
	\nomenclature[Sfc]{$\mathcal{F}^\text{commercial}$}{Commercial flexibility NOE}
	\nomenclature[Ssgl]{$\Delta \mathcal{S}^\lambda$}{set of possible power deviations from the current dispatch point of DERA}
	\nomenclature[D1z]{$\zeta_d$}{Curtailment factor of demand $d$}
	\nomenclature[DPi]{$\boldsymbol{s^{\text{+}}}$}{Power import of DERA; $\boldsymbol{s^{\text{+}}}=[p^{\text{+}},q^{\text{+}}]^T$}
	\nomenclature[Dti]{$t_{ij}$}{OLTC transformer ratio}
	\nomenclature[Pcm]{$c$}{Flexiblity cost}
	\nomenclature[Pgr]{$\boldsymbol{\rho}_r$}{Power deviation price of resource $r$; $\boldsymbol{\rho}_r=[\rho_r^p,\rho_r^q ]^T$}
	\nomenclature[Ptact]{${t}_r^{\text{act-/+}}$}{activation/deactivation time of resources $r$}
	\nomenclature[PtD]{$\Delta t$}{Energy market clearance interval}
	\nomenclature[Pk]{$K$}{Number of equidistant point along reactive power of NOE}
	\nomenclature[Pn]{$N$}{Parameter to control number of points per MVAr of NOE}
	% Note that keywords are not normally used for peerreview papers.
	%\vspace{-0.5em}

	% For peer review papers, you can put extra information on the cover
	% page as needed:
	% \ifCLASSOPTIONpeerreview
	% \begin{center} \bfseries EDICS Category: 3-BBND \end{center}
	% \fi
	%
	% For peerreview papers, this IEEEtran command inserts a page break and
	% creates the second title. It will be ignored for other modes.
	\IEEEpeerreviewmaketitle
	
	%\printnomenclature
	\vspace{-1.5em}
	\section{Introduction}
	%\vspace{-1.0em}
	\bstctlcite{IEEEexample:BSTcontrol}
	\IEEEPARstart{A}{dvancement} in smart grid technologies is enhancing observability and controllability over distributed energy resources (DER), generating opportunity for distribution systems to adopt a more active role and provide operational flexibility to renewables-rich grids~\cite{Capitanescu2018b,Heleno2015,Gonzalez2018,Silva2018}. Various aggregation mechanisms such as virtual power plant (VPP) and transmission and distribution system operator (TSO and DSO) interactions seek to overcome the integration challenges through aggregation of DER and participation in distributed energy marketplaces~\cite{Gonzalez2018,Heleno2015,Silva2018,Wang2019a,Saint-Pierre2016,Xiao2018}. However, despite several works tackling VPP participation in different markets~\cite{Mashhour2011a,Baringo2017}, harnessing multi-energy-vector flexibility~\cite{Giuntoli2013}, proposing active monitoring and real-time control~\cite{DallAnese2017b}, and managing DER uncertainty~\cite{Saint-Pierre2016,Koraki2017,Kardakos2016,Kalantar-Neyestanaki2020}, the actual capacity and constraints of DER aggregations (DERA) to participate in multiple markets has rarely been discussed~\cite{Riaz19}. Furthermore, only a few works, in the context of TSO-DSO interactions, have sought to address the operational ability of DERA while considering \emph{network constraints}~\cite{Capitanescu2018b,Gonzalez2018,Heleno2015,Silva2018,Riaz19,Contreras2018,Noto2019}. This is however essential for all the discussions concerning distributed energy markets, besides TSO/DSO integrated operation. 
	
	Operational flexibility has generally been defined in~\cite{Ulbig2015} as “the technical ability of a component to regulate its power exchange with the grid”. A more detailed definition is presented in~\cite{Zhao2016}, identifying response time and cost as flexibility metrics. Similarly, the capacity-ramp-energy flexibility metric triplet was proposed in~\cite{Ulbig2015,Makarov2009}, but with no reference to cost. Furthermore, the existing literature largely investigates flexibility metrics at the system level, ignoring the impacts of the network~\cite{Ulbig2015,Makarov2009} or reactive power~\cite{Ulbig2015,Makarov2009,Zhao2016}.
	
	Generally speaking, there is no clear definition or even description of the salient features of DERA flexibility, including its steady-state and dynamic aspects and how these are impacted by network constraints and market requirements. The impact on DERA market participation from limited ramp rate~\cite{Riaz19} and energy availability~\cite{Wang2019a} may for example be significant. It is in this context that it may be beneficial to recognise the distinct role of “steady-state” flexibility and “dynamic” aspects of flexibility, where \emph{steady-state} flexibility indicates the possibility to operate within a certain operational envelope regardless of temporal factors,~\cite{Riaz19}, while \emph{dynamic} flexibility also accounts for time-varying aspects. This could include \emph{notice time} to deliver flexibility, minimum time to reach a new operating point (\emph{speed} to change the operating state, in case already including the notice time too), the maximum \emph{duration} for which the new operating point can be sustained, etc.~\cite{Riaz19,Mancarella2013C}. However, despite their key importance, these dynamic aspects of flexibility are only sometimes considered in studies focused at the system level~\cite{Nosair2015,Lannoye2015a} and only recently for works modelling distribution level flexibility~\cite{Stanojev2021,Gasser2021}.
	
	In line with but also expanding general system flexibility concepts, we can then identify \emph{capacity}, \emph{ramp}, \emph{duration} and \emph{cost} as key metrics to assess DERA flexibility. Furthermore, flexibility assessment at the distribution level calls for explicit consideration of reactive power and network constraints, which can be evaluated and visualised in active-reactive (P-Q) space~\cite{Capitanescu2018b,Gonzalez2018,Heleno2015,Silva2018,Riaz19,Contreras2018,Noto2019,Wang2019a,Liu2021}. The concept of P-Q charts was first developed in~\cite{Cuffe2014} to represent power exchange between TSO and DSO, and different applications have been proposed, including a P-Q chart-based planning and operational methodology to deal with uncertainty for reserve provision~\cite{Kalantar-Neyestanaki2020}. However, all of these works only focused on the capacity metric, ignoring dynamics and cost aspects of flexibility. Dynamic aspects of flexibility are also ignored by~\cite{Silva2018,Heleno2015} when building flexibility P-Q cost maps as an effective information exchange tool between TSO and DSO.
	
	On the above premises, and bearing in mind the distinction between virtual/physical (i.e., without/with consideration for the presence of the network) aggregation introduced in the FENIX project~\cite{Pudjianto2007}, this paper proposes a comprehensive framework to model, describe, characterise and quantify the flexibility that could be provided upstream by a DERA considering all relevant flexibility metrics, technical and commercial aspects, steady-state and dynamic characteristics, and network restrictions. To do so, we introduce the concept of \emph{nodal operating envelope} (NOE) that can be visualised through active-reactive power diagrams in a distribution network. Different yet consistent NOEs can be defined to highlight and distinguish \emph{key features} associated with DERA flexibility, namely: i) aggregated capacity from virtual clustering of DER (\emph{capability}); ii) role of the local network in enabling and constraining DER flexibility (\emph{feasibility}); iii) dynamic ability to move between two points within the feasibility region (\emph{ramp flexibility}); iv) energy capability to sustain an operating point for the required call length of a specific service (\emph{duration flexibility}); v) cost associated with changing operating points and provide a certain service (\emph{economic flexibility}); vi) technical ability to provide a service (\emph{technical flexibility}); and vii) actual market participation potential from the techno-economic perspectives (\emph{commercial flexibility}). In particular, the quantification of dynamic aspects of flexibility serves as a cornerstone to assess the potential for DERA to partake in energy and grid support services, thus enhancing their visibility and supporting DER business cases.
	
	The specific contributions of this paper are as follows:
	\vspace{-.25em}	
	\begin{itemize}[leftmargin=*]
		\item Identify capacity, ramp, duration and cost as key metrics to assess DERA flexibility;
		\item Definition and classification of salient features of DERA \emph{steady-state} and \emph{dynamic} flexibility (i.e., capability, feasibility, ramp, duration, economic, technical and commercial);
		\item Mathematical description of these features starting from a Venn diagram to provide comprehensive characterisation and consistent terminology for DER flexibility analysis;
		\item Systematic representation of the different flexibility features at different aggregation levels (“nodes”) via \emph{nodal operating envelopes}, visualised in “familiar” active-reactive power space and illustrated via a canonical example;
		\item Extension of optimal power flow analysis (OPF), only presented in the literature for feasibility, to quantify all the proposed DER flexibility features, while explicitly considering operational and network limits, cost due to DER flexibility activation, and market requirements;
		\item Representation of different flexibility features as linear constraints to ensure scalability of the approach;
		\item Bid stack creation of DERA for multi-market participation;
		\item Case studies from a canonical and a real Australian distribution network demonstrating the efficacy of the proposed framework in modelling the various DER flexibility features. 	 
	\end{itemize}
	
	The rest of the paper is organised as follows: Section~\ref{Sec:Key Def} outlines key fundamental concepts and provides a canonical illustrative example. The mathematical model and the methodology to determine flexibility NOEs are presented in Section~\ref{Sec:NOE framework}. The case studies from a real network are discussed in Section~\ref{Sec:Case Study}. Finally, Section~\ref{Sec:Conclusion} concludes the paper.
	
	\vspace{-0.25em}
	\section{Key Concepts and Definition} \label{Sec:Key Def}
	%\vspace{-0.25em}
	\subsection{Nodal operating envelope (NOE)}
	\vspace{-0.25em}
	An \emph{operating envelope} essentially represents the operating region of a device or system as described by relevant constraints. Such an operating envelope, which depends on several parameters and variables, can be then visualised in different spaces that are the projection of the more general multi-dimensional operational region. For example, in the simplest one-dimension (1-D) form, envelopes can be described as a function of active power only. Moving to higher dimensions, a 2-D envelope can represent the operating space as a function of active power in conjuncture with one other parameter such as time, variability, voltage, ramp, duration, cost, reactive power, or even other energy vectors (e.g., heat, gas, etc.)~\cite{Chicco2020}, and so forth. The choice of the sub-space to represent the envelope may depend upon the use case. For example, at the system level (where reactive power does not play a role as important as in distribution networks), the relevant operational envelopes of the region under study could be represented as a function of active power only (DC power flow approximation) and against wind variability and with respect to time, as for example in ~\cite{Nosair2015}. Another alternative operating envelope representation could be in capacity-ramp-duration space, as for instance in ~\cite{Ulbig2015,Makarov2009}.
	
	In this work we aim to determine the DERA flexibility across different network location and voltage levels, thus we introduce the concept of NOE that represents the aggregated operating envelope of the downstream DER (with respect to a reference \emph{aggregation point}). This reference aggregation point (“\emph{node}”) can be \emph{virtual} (i.e., not considering the physical presence and in case constraints of a network) or \emph{physical} (as for specific network nodes and relevant downstream constraints). Hence, we talk about \emph{nodal} envelopes as a means to reflect the aggregated behaviour of the downstream DER with respect to a reference node.
			
	In the presented work, NOEs are mathematically defined via set notation or using a set of linear constraints of the form $\boldsymbol{A}\boldsymbol{x}\leq\boldsymbol{b}$. The linear constraint representation allows more convenient incorporation of the NOE in market clearance engines, as further explained in Section~\ref{Sec:Results}. On the other hand, the set notation enables a compact representation of NOEs at a conceptual level and as used in Section~\ref{Sec:FLC_xtics}. Furthermore, we have extensively visualised NOEs in the P-Q Euclidean space (in which P-Q charts are drawn) due to their familiarity in the power systems. With this representation, the special case of capability/feasibility NOE coincides with the familiar P-Q chart. However, while this is only a static concept, the concept of flexibility NOE is more general as it can also represent, in a synthetic way ramp-rate, energy content, cost of DERA, and potentially other features too, besides the impact of network constraints and market requirements.
	
	In summary, the NOEs are effectively a mathematical formulation of the operational space and flexibility potential of the system considered downstream a given node. In addition, Section~\ref{Sec:NOE_UseCase} discusses the use of NOEs to quantify different flexibility features, and Section~\ref{Sec:Scalability NOE} demonstrates how the concept of NOE is flexible and scalable: the reference point can be suitably chosen and flexibly moved across different aggregation levels depending on the specific study.
	\vspace{-1.0em}
	\subsection{Flexibility characterisation and definitions } \label{Sec:FLC_xtics}
\begin{figure}[]
			\centering
			\includegraphics[width=83mm]{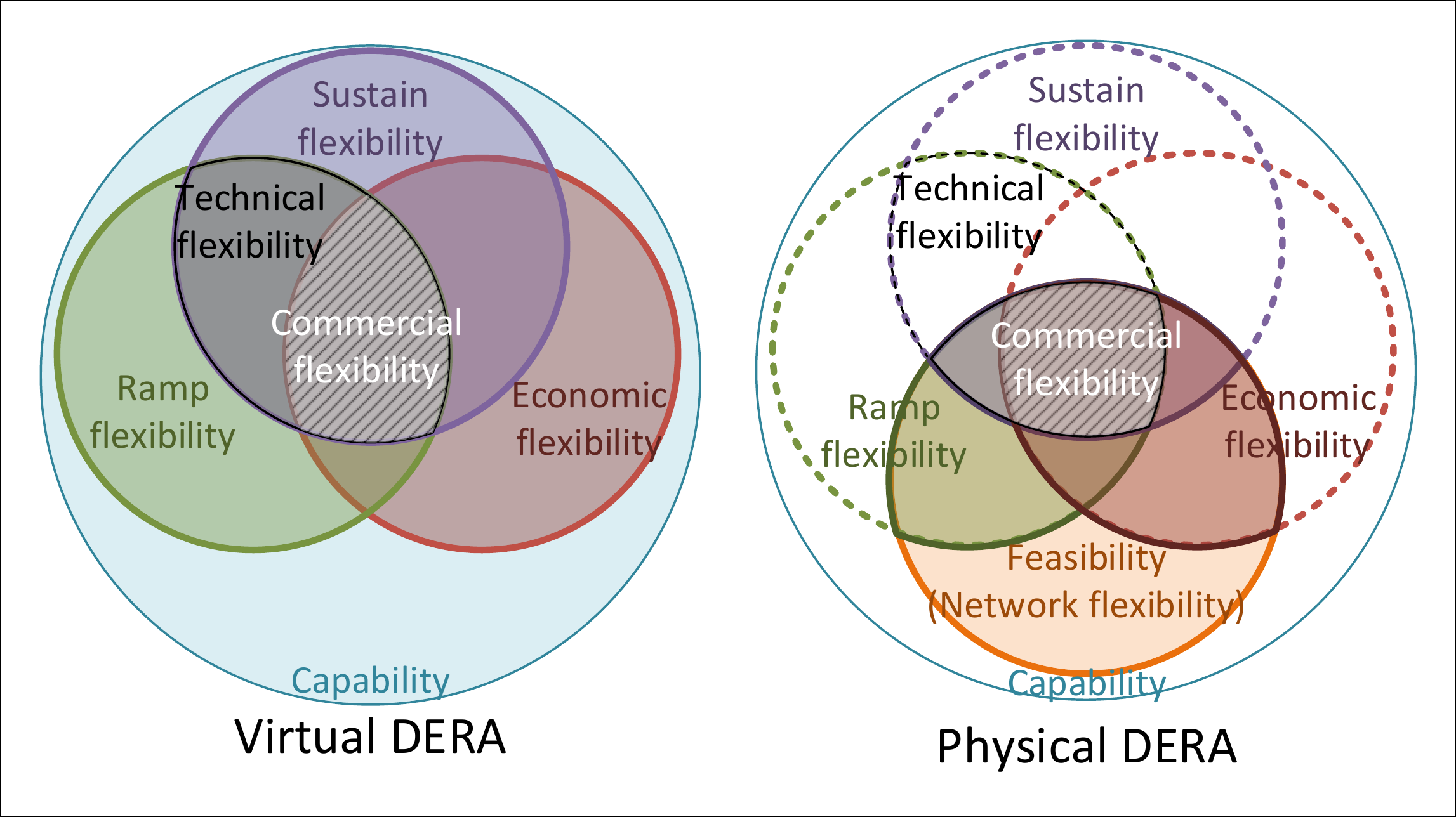}
			\vspace{-0.5em}
			\caption{\emph{Venn diagram} showing the relationship between virtual (left) and physical (right) DERA flexibility features. The dotted lines represent the flexibility NOEs that cannot be deployed because of network constraint.}
			\vspace{-1.0em}
			\label{fig:Flx_VenDia}
		\end{figure}

	\vspace{-0.25em}
	As discussed in~\cite{Gonzalez2018,Riaz19}, there is an emerging need for characterising and defining flexibility, also using a more consistent terminology in the context of DERA. In fact, DERA flexibility can be intuitively associated with the ability of the aggregate to alter its power exchange with the grid in a given time interval, and expressed in terms of the \emph{capacity-ramp-duration-cost quadruplet} that assesses DERA flexibility potential independently of any service requirement. The relationship among the flexibility features introduced here for a virtual and physical DERA are summarised in Fig.~\ref{fig:Flx_VenDia}.
		In the figure, capability and feasibility represent the “capacity” metrics for a virtual and physical aggregation, respectively. For a given DERA, ramp flexibility and duration flexibility refer to the “ramp” and “duration” metrics, respectively, while economic flexibility refers to the “cost” metric. Furthermore, the concepts of technical and commercial flexibility (derived from the flexibility quadruplet but also considering specific service requirements) are introduced to demonstrate a DERA ability to participate in a specific market segment from a technical and techno-economic perspective, respectively.
	
	The different flexibility features are further discussed below based on a canonical example.
	
	%\vspace{-1.25em}
	\subsubsection{Illustrative canonical example} \label{Sec:Exp}
\begin{figure}[]
			\centering
			\includegraphics[width=70mm] {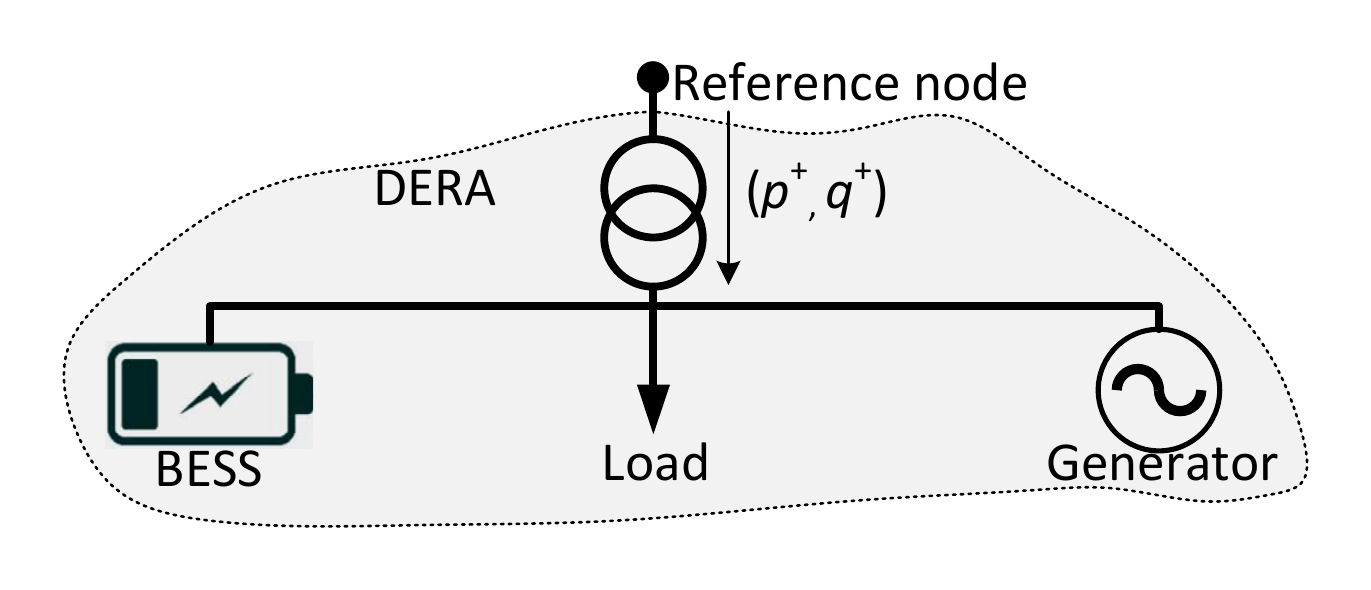}
			\vspace{-1.0em}
			\caption{Single line diagram of the DERA of the canonical example.}
			\vspace{-1.5em}
			\label{fig:Exp_Net}
		\end{figure}

		Fig.~\ref{fig:Exp_Net}
		shows a simple example to illustrate the DERA flexibility features and relevant NOEs, with a diesel generator, battery energy storage system (BESS), and load of \SI{0.47}{\mega \watt} and \SI{0.22}{\mega \var}. The reference node (for NOE) is taken at the primary of the transformer. The transformer’s thermal limit is taken as \SI{1.14}{\mega \voltampere}. The BESS is assumed to be charged at \SI{2.1}{\percent} of its maximum energy value, and both the generator and BESS are dispatched at zero active and reactive power. Other key parameters are shown in Table~\ref{Table:DER_Param}.
		\vspace{-0.5em}
	\begin{table}[t]
		% increase table row spacing, adjust to taste
		\renewcommand{\arraystretch}{1.3}
		\caption{Characteristics of resources considered for the test DERA in Section~\ref{Sec:Exp}.}
		\label{Table:DER_Param}
		\centering
		\vspace{-1.0em}
		\begin{tabular}{|l|c|c|c|c|}
			\hline
			Resource&Rated power&Activation&Ramp rate&Deviation cost\\ 
			&[\SI{}{\mega \watt}]&delay\footnote{} [\SI{}{\second}]&[\SI{}{\mega \watt\per\second}]&[\SI{}{\$\per\mega\watt\hour}]\\ \hline
			Generator&\SI{1}{}&\SI{25}{}&\SI{0.033}{}&\SI{380}{}\\ \hline 	
			BESS&\SI{0.5}{}&\SI{0.5}{}&\SI{1.67}{}&\SI{190}{}\\ \hline
		\end{tabular}
		\vspace{-0.5em}
	\end{table}
 	\vspace{-0.5em}
	\subsubsection{Capability} 
		For a given time interval and in a generic \emph{virtual} aggregation, the set of viable points from aggregating the individual active-reactive (P-Q) operating regions represents what we call here \emph{capability} ($\mathcal{C}$) of the DERA. Mathematically, the relevant capability NOE can be estimated by performing Minkowski summation over the individual DER P-Q regions:
	\vspace{-0.50em}
	\begin{equation} \label{Eq:Capability}
		\mathcal{C}=\{\sum_{r\in \mathcal{R}}\boldsymbol{s}_r:\boldsymbol{s}_r\in \mathcal{S}_r\} \text{,}
		\vspace{-.5em}
		\end{equation}
		where, $\mathcal{R}$ is the set of all resources, $\mathcal{S}_r$ represents the P-Q capability set of resource $r$ and $\boldsymbol{s}_r=[p_r,q_r ]^T$ is a single operating point within individual DER capability set $\mathcal{S}_r$. Furthermore, in a virtual DERA the dispatch point ($\boldsymbol{s}^\lambda=[p^\lambda,q^\lambda ]^T$) is independent of the network and therefore can be mathematically described as $\boldsymbol{s}^\lambda=\sum_{r \in \mathcal{R}} \boldsymbol{s}_r$.
	
	\footnotetext[1]{Time required to activate the resource including communication delay.}
	
\begin{figure}[]
			\centering
			\includegraphics[width=87mm] {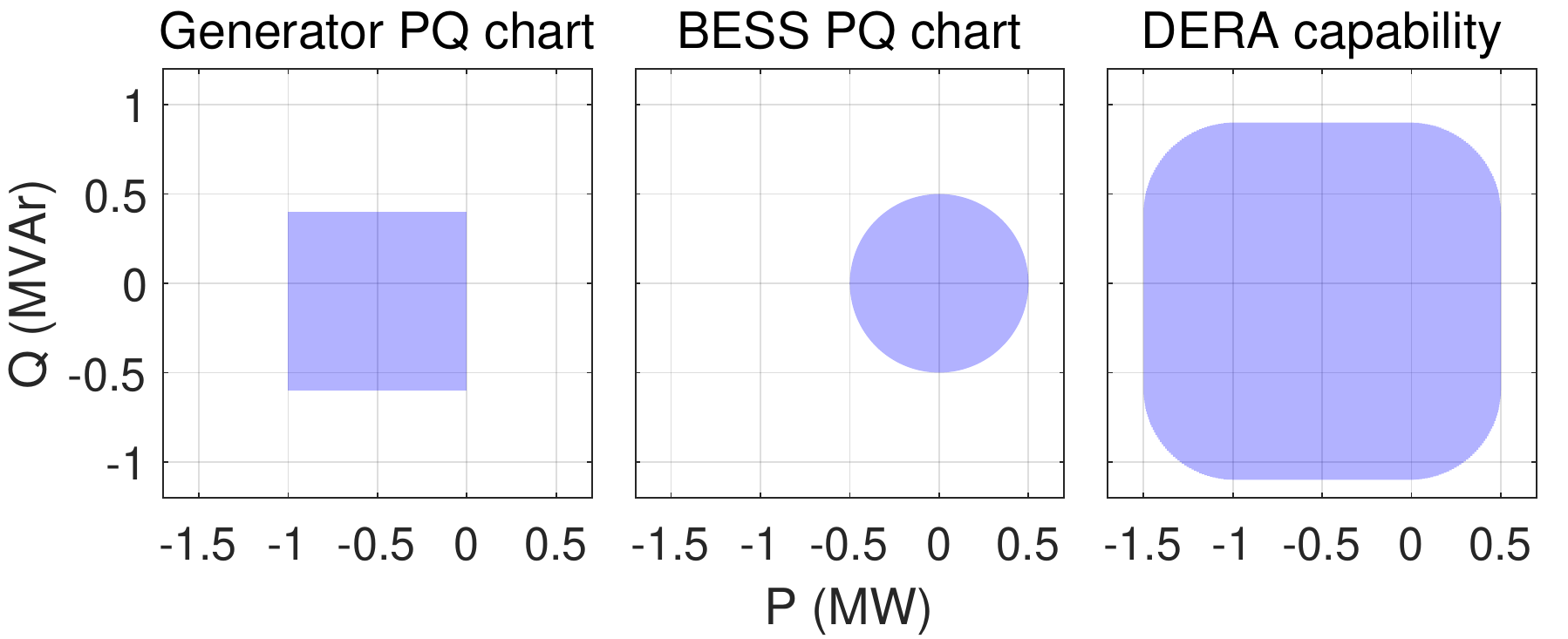}
			\vspace{-1.0em}
			\caption{Individual P-Q capability charts of generator (left) and BESS (middle), along with the DERA \emph{capability} NOE (right).}
			\vspace{-1.5em}
			\label{fig:Exp_Cap}
		\end{figure}

	The P-Q charts of generator and BESS along with the DERA \emph{capability} NOE (obtained by Minkowski summation) are shown in Fig.~\ref{fig:Exp_Cap}.
		The BESS output is limited by the inverter rating, while the synchronous generator P-Q chart is depicted in a simplified way by restricting the operational range of reactive power, as normal in such studies~\cite{Capitanescu2018b,Silva2018,Riaz19} and especially acceptable for illustration purposes. Note that the aggregation of the two DER results in a much larger NOE, which underscores the benefits derived from aggregation.
	\subsubsection{Feasibility}
		With deeper penetration of DER, the set of viable operating points, will increasingly be limited by network constraints. The resulting NOE can therefore be associated with the concept of \emph{feasibility}, in the sense that the potential operating points now must also be network-feasible. Feasibility ($\mathcal{F}$) can therefore be defined as the set of all the network feasible P-Q dispatch points of a DERA, where each dispatch point $\boldsymbol{s}^\lambda=f(\boldsymbol{Y},\boldsymbol{s}_d,\boldsymbol{x})$ is now influenced by grid parameters ($\boldsymbol{Y}$), load requirement ($\boldsymbol{s}_d=[p_d,q_d ]^T$) and control variables ($\boldsymbol{x}$):
	\vspace{-0.25em}
	\begin{equation} \label{Eq:Feasibility}
		\mathcal{F}=\{\mathcal{C}+\sum_{d \in \mathcal{D}} \boldsymbol{s}_d + \sum_{l \in \mathcal{L}} \boldsymbol{s}_l^{\text{loss}} : \underline{\boldsymbol{v}_i} \leq \boldsymbol{v}_i \leq \overline{{\boldsymbol{v}_i}} \land  |\boldsymbol{s_l}|\leq \overline{\boldsymbol{s}} \} \text{,}
		\vspace{-0.25em} 
		\end{equation}
		where, $\mathcal{L}$ denotes the set of network elements (lines, transformers, etc.), $\boldsymbol{s}_l^{\text{loss}}$ and $\boldsymbol{s_l}=[p_l,q_l ]^T$ are power loss and power flow in element $l$, respectively. Furthermore, $\overline{\boldsymbol{v}}_i$/$\underline{\boldsymbol{v}}_i$ represent the upper/lower bound of the bus $i$ voltage $\boldsymbol{v}_i= v_i\angle\theta_i$.

\begin{figure}[]
			\centering
			\includegraphics[width=\linewidth] {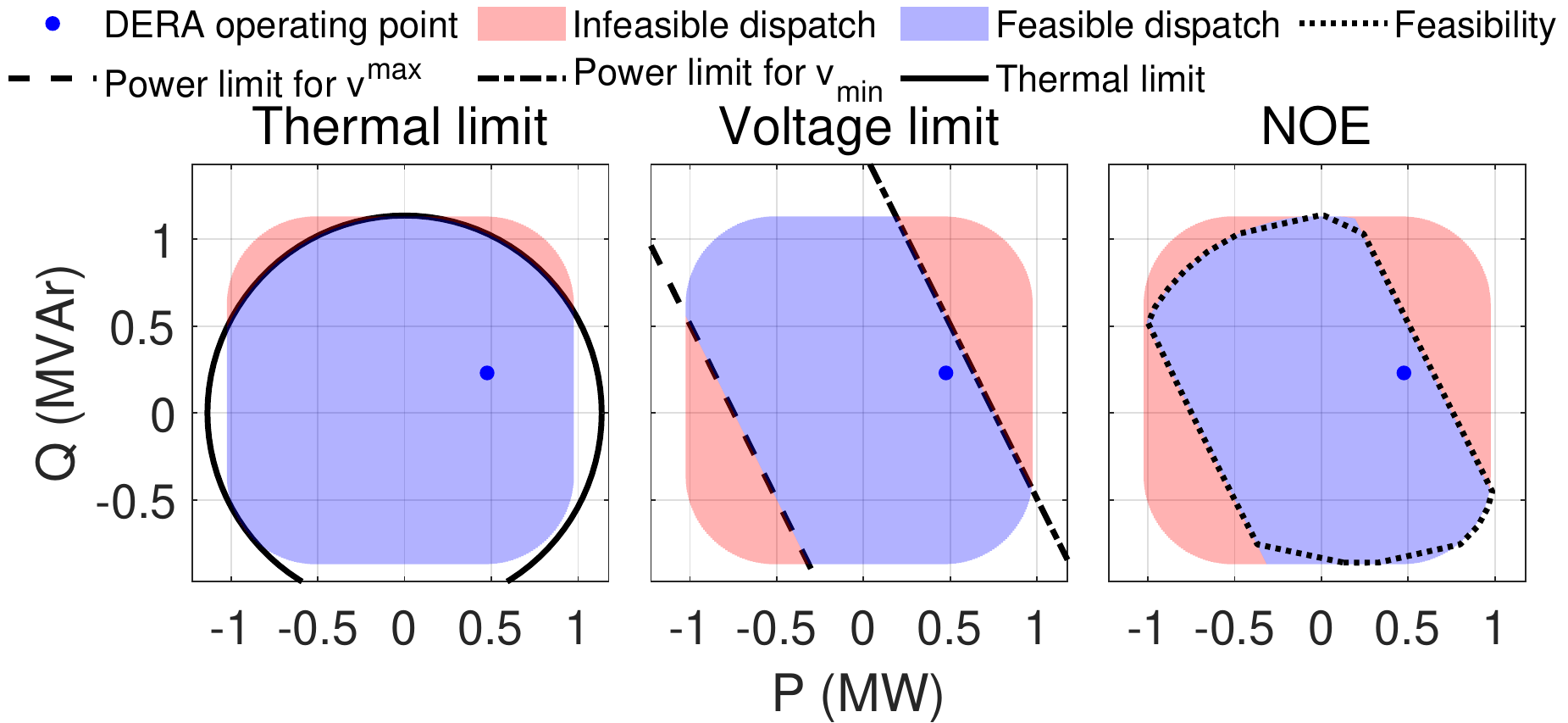}
			\vspace{-2em}
			\caption{Illustration of thermal (left) and voltage (middle) network constraints, with the resultant \emph{feasibility} NOE (right) of the test DERA.}
			\vspace{-0.75em}
			\label{fig:Exp_FOR}
		\end{figure}

	The capability envelope provides information on the operational range of virtual aggregation, and the operation of DER can be modelled \emph{independently} of the demand requirement. However, in a physical aggregation setup the network is shared by loads and DER and introduces physical restrictions (e.g., thermal and voltage limits). The effect of network is shown in Fig.~\ref{fig:Exp_FOR},
		representing the impact of thermal (left) and voltage (middle) constraints and the DERA \emph{feasibility} NOE (right). 
	\subsubsection{Ramp and duration (dynamic) flexibility}
		While many papers associate DERA flexibility with what we have defined as capability and feasibility~\cite{Gonzalez2018,Heleno2015,Silva2018,Contreras2018}, in reality these lack critical information regarding dynamic aspects of service provision. Thus, we propose the following.
	
	\emph{Ramp flexibility} quantifies the operational envelope around a given dispatch point with respect to DERA response time $\tau$: this depends on the up/down ramp capability ($r_r^{\text{+/-}}$), commitment status, and active power dispatch point ($p_r$) of DER. Similarly, the \emph{duration flexibility} measures the operational envelope with respect to the duration $\psi$ for which the response can be sustained: this depends upon the DER energy content ($e_r$). The relevant NOEs for ramp and duration flexibility can be defined as:
	%\vspace{-0.25em}
	\begin{equation} \label{Eq:RampNOE}
		\mathcal{F}^\text{ramp}=\{\Delta \mathcal{S}^\lambda:\Delta p_r/r^{\text{+/-}}\leq \tau,\forall r \in \mathcal{R}\} \text{,}
		%\vspace{-0.25em}
	\end{equation}	
	\begin{equation} \label{Eq:DurationNOE}
		\mathcal{F}^\text{duration}=\{\Delta \mathcal{S}^\lambda:e_r/\Delta p_r\geq \psi,\forall r \in \mathcal{R}\} \text{,}
		%\vspace{-0.25em}
	\end{equation}
	where $\Delta \mathcal{S}^\lambda$ is the set of possible power deviations from the current dispatch point of DERA, which (in virtual aggregation) can be defined as {$\Delta \mathcal{S}^\lambda=\mathcal{C}-\boldsymbol{s}^\lambda$, and $\Delta \mathcal{S}^\lambda = \mathcal{F} - \boldsymbol{s}^\lambda - \sum_{l \in \mathcal{L}}\Delta \boldsymbol{s}_l^\text{loss}$} for physical aggregation, while $\Delta p_r$ representing the active power deviation of individual DER.
	
	\begin{figure}[]
		\centering
		%\vspace{-1.5em}
		\includegraphics[width=84mm] {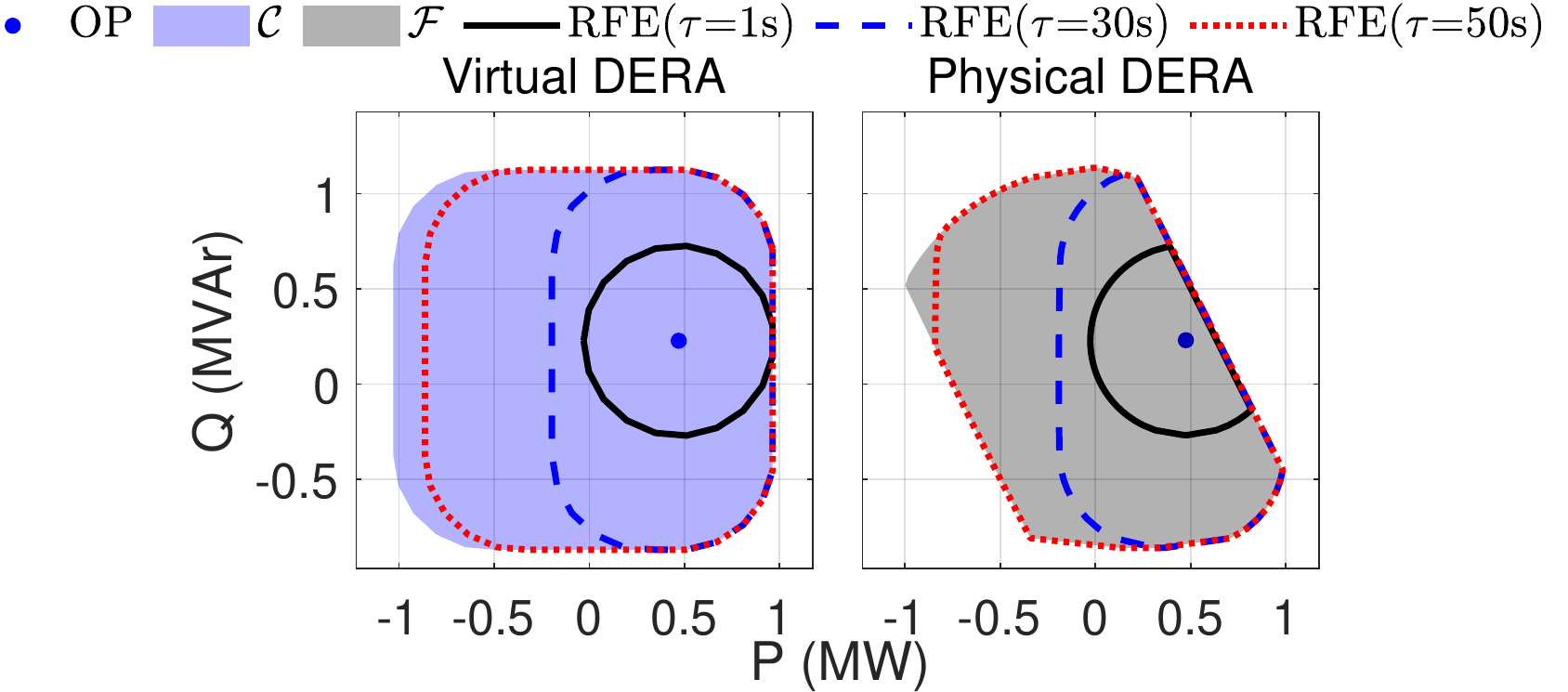}
		\vspace{-1.0em}
		\caption{\emph{Iso-ramp} contours of \emph{ramp flexibility} NOE (RFE) of the test DERA under virtual (left) and physical (right) arrangements.}
		\vspace{-1.75em}
		\label{fig:Exp_RFLX}
	\end{figure}	
	
\begin{figure}[]
			\centering
			\includegraphics[width=84mm] {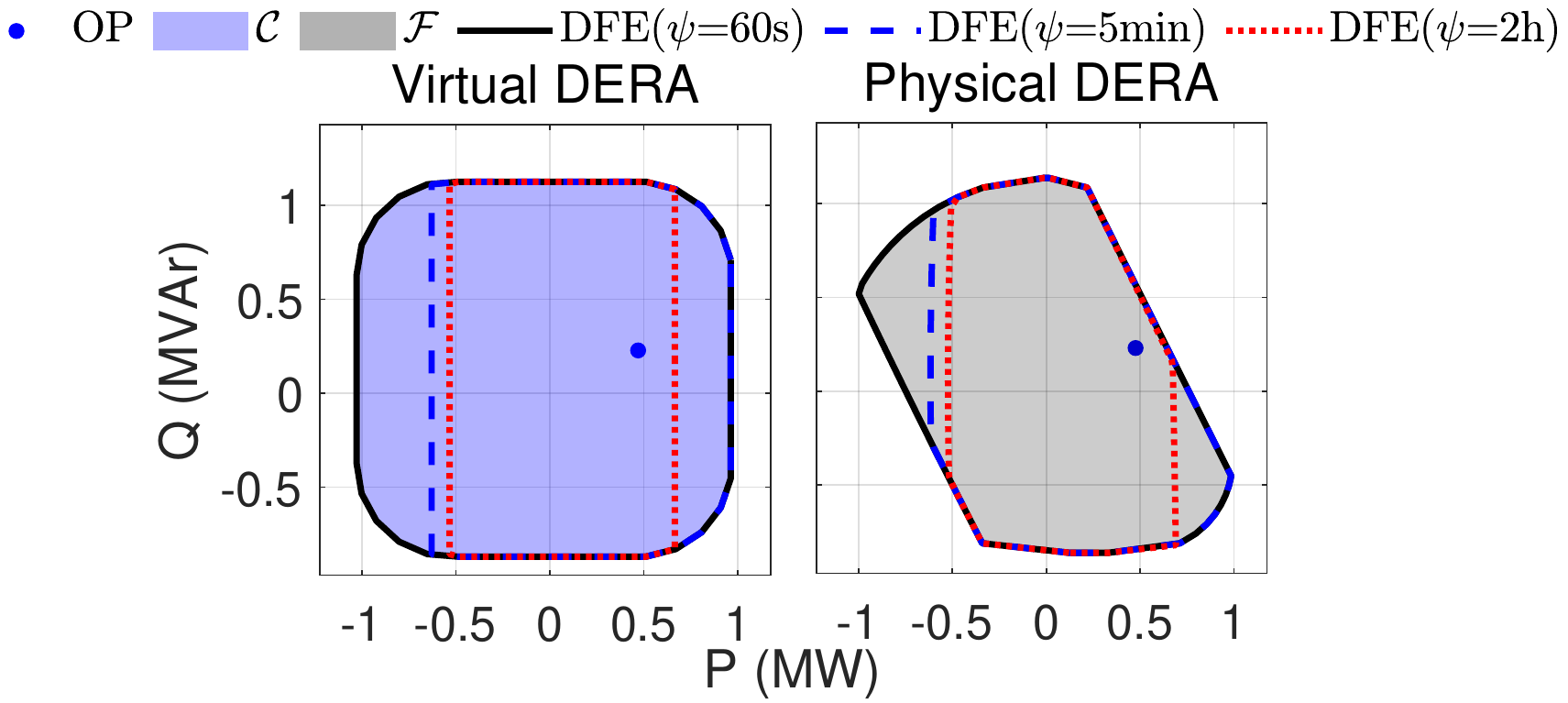}
			\vspace{-0.75em}
			\caption{\emph{Iso-duration} contours of \emph{duration flexibility} NOE (DFE) of the test DERA under virtual (left) and physical (right) arrangements.}
			\vspace{-0.75em}
			\label{fig:Exp_DFLX}
		\end{figure}

	The role of ramp rate and energy content of DERA can be visualised through the \emph{ramp} and \emph{duration} flexibility NOEs (RFE and DFE), respectively. Both RFE and DFE not only depend upon the active and reactive power but also on the service response time ($\tau$) and duration ($\psi$), respectively. Thus, the RFE and DFE results in 3D volumes as illustrated in Appendix-I. However, due to difficulty in representing 3D volumes on 2D surfaces, instead, we represent the \emph{iso-ramp} and \emph{iso-duration} contour of RFE (Fig.~\ref{fig:Exp_RFLX}) and DFE (Fig.~\ref{fig:Exp_DFLX}) on the P-Q plane, respectively. The choice of the P-Q plane is because its familiarity in the power system and because each service has a specific response time and duration, thus resulting in at most one contour per service.

	Fig.~\ref{fig:Exp_RFLX} quantifies the DERA potential to deviate from the current operating point (OP) within a certain response time ($\tau$) in a virtual (left) and physical (right) aggregation. The size of RFE monotonically \emph{increases} with response time, as more resources can be deployed (ramp up/down) to deliver the required response. Similarly, the DFE under virtual (left) and physical (right) arrangement are shown in Fig.~\ref{fig:Exp_DFLX}.
		%\vspace{0em}
		In contrast to RFE, the DFE monotonically \emph{decreases} with service length, as energy-constrained resources cannot maintain the same power deviation for longer durations.	
	
	\subsubsection{Economic flexibility} 
		Deviation from a certain dispatch point will incur some cost, which is associated with the concept of \emph{economic flexibility} and mathematically defined as:
	\vspace{-0.25em}
	\begin{equation} \label{Eq:EconomicNOE}
		\mathcal{F}^\text{economic}=\{\Delta \mathcal{S}^\lambda:\sum_{r\in\mathcal{R}}\boldsymbol{\rho}_r^T |\Delta \boldsymbol{s}_r| \leq c\} \text{,}
		\vspace{-0.5em}
		\end{equation}
		where $\Delta \boldsymbol{s}_r=[\Delta p_r,\Delta q_r ]^T$, $\boldsymbol{\rho}_r=[\rho_r^p,\rho_r^q ]^T$, and $c$ represent the power deviation and deviation price of individual DER, and the cost the DERA is willing to incur, respectively. The power deviation cost $\boldsymbol{\rho}_r$ is attributed to all the relevant factors such as additional cost of fuel, variable maintenance (e.g., ramping fatigue), lost opportunity (active power being restricted due to provision of reactive power), electricity buy back in case of storage, etc.

\begin{figure}[]
			\centering
			%\vspace{-1em}
			\includegraphics[width=\linewidth] {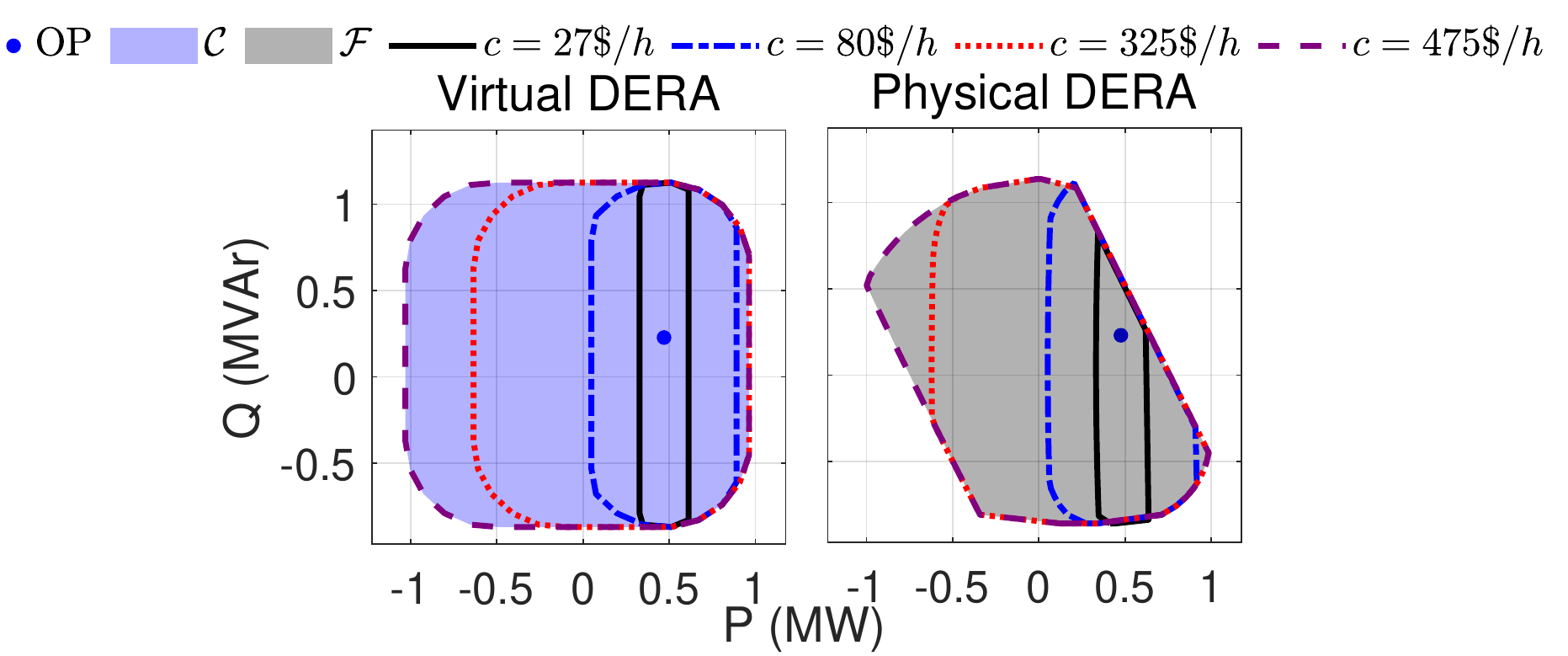}
			\vspace{-1.75em}
			\caption{\emph{Iso-cost} curves of the test DERA representing \emph{economic flexibility} NOE under virtual (left) and physical (right) arrangements.}
			\vspace{-1.5em}
			\label{fig:Exp_EFLX}
		\end{figure}

	The cost to deviate from the current dispatch point also has a significant influence on the DERA potential to partake in various markets. The 3D \emph{economic flexibility} NOE (EFE) of the test DERA is represented via \emph{iso-cost} curves in Fig.~\ref{fig:Exp_EFLX}
		(see Appendix-I for 3D representation of the EFE), where the \emph{iso-cost} curves enclose the operational envelope of the DERA for a given cost $c$. The EFE reflect the financial willingness of DER to deviate from their dispatch point and show a proportional relationship with $c$.

	\subsubsection{Technical and commercial (market specific) flexibility} Given a certain DERA portfolio, the capability/feasibility, ramp, duration and economic NOEs reflect its flexibility potential, independently of any particular service requirement. However, building on the above discussions, it may be advantageous to identify the flexibility features that quantify the ability of a DERA to participate in a particular service/market segment by considering the market-specific techno-economic requirements or arrangements, defined below as technical and commercial flexibility. More specifically:
	
	\emph{Technical flexibility} can be defined as the ability of a DERA to deviate from the current dispatch point considering market/service-specific technical constraints. For simplicity, we only consider response time ($\tau$) and call length ($\psi$) requirements; however, for a specific service all the relevant constraints should be included. In our general and generic case the technical flexibility NOE can be calculated as:
	%\vspace{-0.25em}
	\begin{equation}
		\mathcal{F}^\text{technical}=\{\Delta \mathcal{S}^\lambda:\Delta p_r/r^{\text{+/-}}\leq \tau \wedge e_r/\Delta p_r\geq {\psi}, \forall r \in \mathcal{R}\} \text{.}
		%\vspace{-0.25em}
		\end{equation}
	
\begin{figure}[]
		\centering
		\vspace{-0.5em}
		\includegraphics[width=61mm] {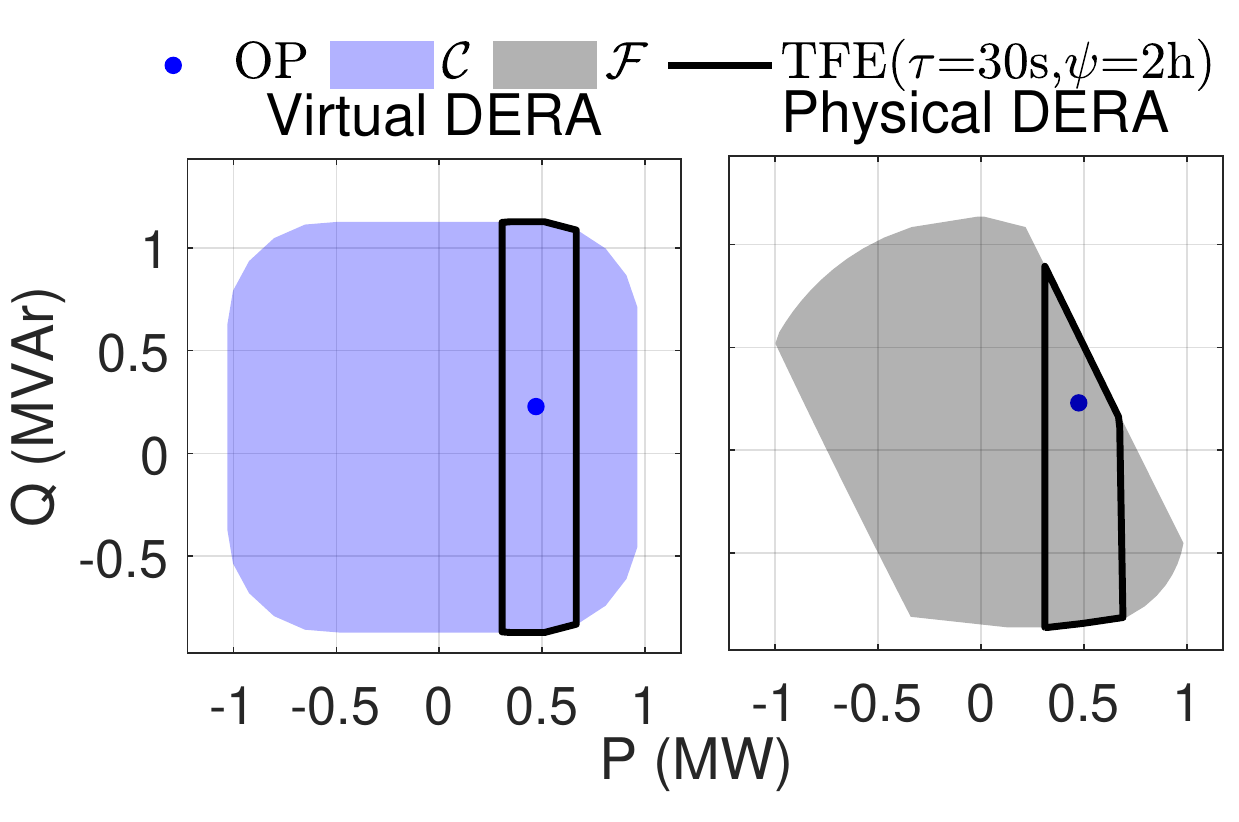}
		\vspace{-0.85em}
		\caption{\emph{Technical flexibility} NOE (TFE) contour (\SI{2}{\hour} service with \SI{30}{\second} response time) of the test DERA for virtual (left) and physical (right) arrangement.}
		\vspace{-1.5em}
		\label{fig:Exp_TFLX}
	\end{figure}

	Therefore, the \emph{technical flexibility} NOE (TFE) can be represented in the active power, reactive power, response time and duration 4D hyperspace. For example, suppose a market/network service is such that response is required within \SI{30}{\second} of an event and for \SI{2}{\hour}. The TFE contour for such service for virtual (left) and physical (right) aggregation is exemplified in Fig.~\ref{fig:Exp_TFLX}.

%\vspace{-0.25em}
	\emph{Commercial flexibility} can be viewed as the ability of a DERA to partake in a market segment considering techno-economic constraints. That is, the DERA commercial flexibility depends on its potential to participate in a market given the required response time ($\tau$), service duration ($\psi$) and anticipated clearance price ($c$):
	\vspace{-0.25em}
	\begin{equation}
		\mathcal{F}^\text{commercial}=\{\mathcal{F}^\text{technical}(\tau,{\psi}):\sum_{r\in\mathcal{R}}\boldsymbol{\rho}_r^T |\Delta \boldsymbol{s}_r| \leq c\} \text{.}
		\vspace{-0.25em}
		\end{equation}
		
	\begin{figure}[]
		\centering
		\includegraphics[width=61mm] {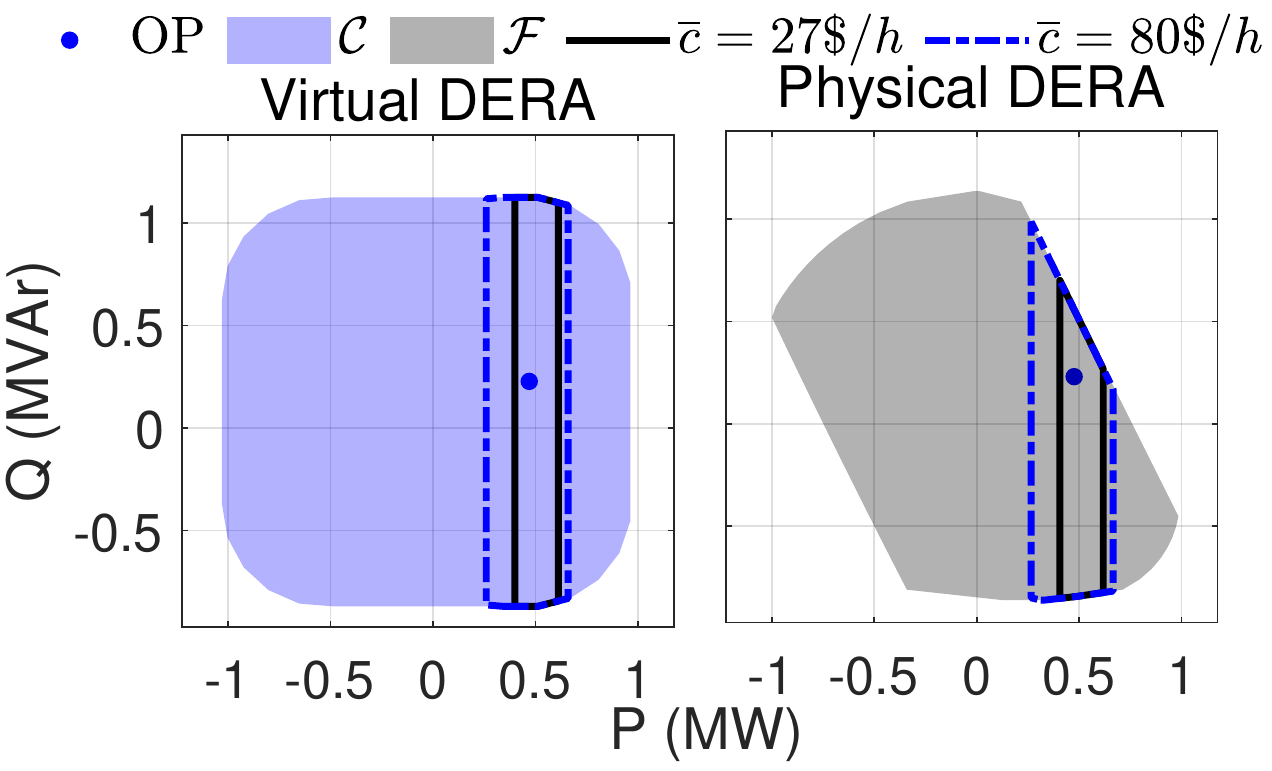}
		\vspace{-0.5em}
		\caption{\emph{Commercial flexibility} NOE (CFE) \emph{iso-cost} contours of the test DERA under virtual (left) and physical (right) aggregation; \SI{2}{\hour} service with \SI{30}{\second} response time.}
		\vspace{-1.0em}
		\label{fig:Exp_CFLX}
	\end{figure}

	Therefore, the \emph{commercial flexibility} NOE (CFE) quantifies the DERA ability to participate in a specific market considering both technical and economic restrictions and can be described in the active power, reactive power, response time, duration and cost 5D hyperspace. The \emph{iso-cost} contours of CFE for the same example as in Fig.~\ref{fig:Exp_TFLX} is shown in Fig.~\ref{fig:Exp_CFLX}, for virtual (left) and physical (right) arrangements, and various values of the anticipated market clearance prices $c$. These CFEs can be thought of as implicit means to reflect the short-run marginal cost of the DERA for a particular service.

	\vspace{-1.25em}
	\subsection{Flexibility features and use cases}\label{Sec:NOE_UseCase}
\begin{figure}[]
		\centering
		\includegraphics[width=\linewidth] {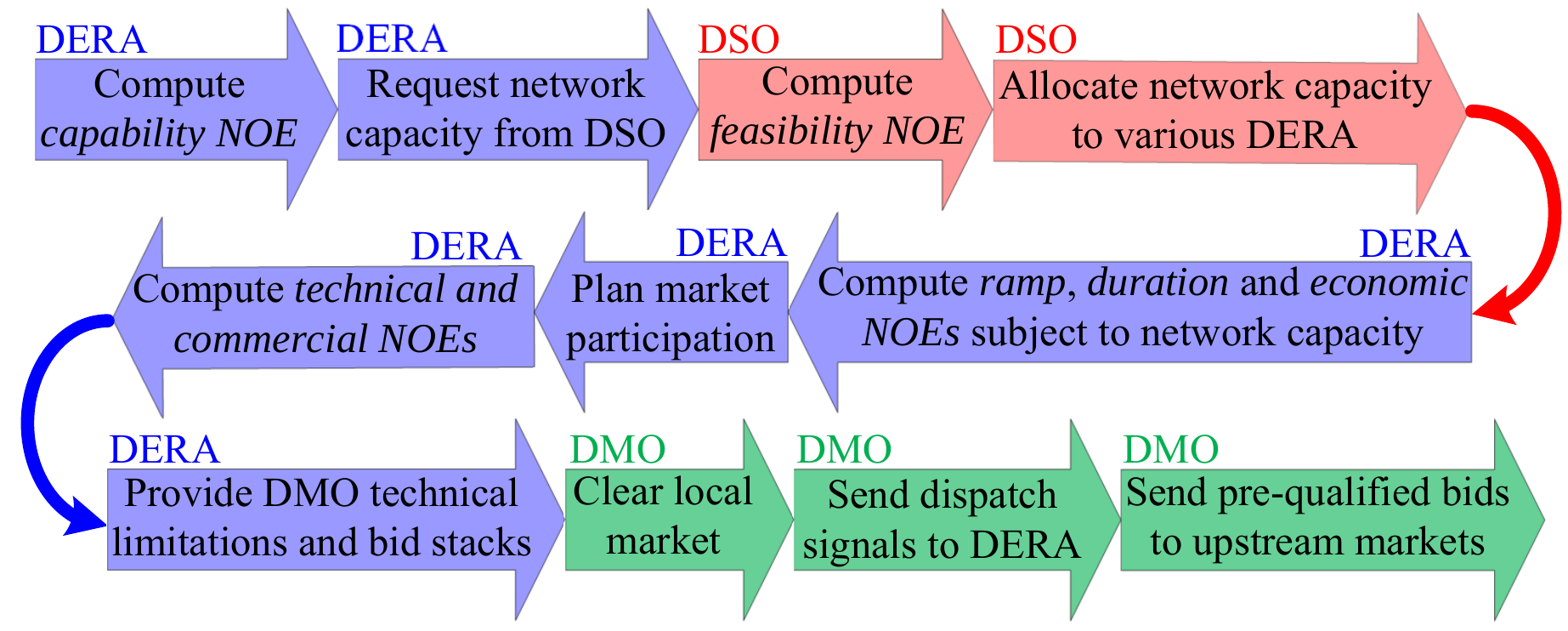}		
		\vspace{-1.75em}
		\caption{A potential use case of various NOEs in a distributed marketplace environment.}
		\vspace{-1.5em}
		\label{fig:Flx_FlChart}
	\end{figure}

	\vspace{-0.25em}
	Different flexibility features may be envisioned to be used by different entities, and the proposed concepts are being practically adopted in Australia to support the development of a distributed energy marketplace~\cite{ARENA2020}. Different technical and commercial mechanisms for interactions between the relevant stakeholders and exchange of NOE information could be proposed, which would be case-dependent. One such case is shown in Fig.~\ref{fig:Flx_FlChart}.
	For example, for each market clearance interval, aggregators could determine their specific DERA portfolio’s \emph{capability} NOE to establish their active and reactive power import and export potential (i.e., a “capacity” metric), and send this information to the DSO for network capacity allocation. The DSO could then aggregate all capability NOEs received from different aggregators and determine the network \emph{feasibility} NOE, which would be used to allocate network capacity to various aggregators. This could also be interpreted as the time-varying hosting capacity of the network (see also Section~\ref{Sec:Scalability NOE}). Aggregators could then perform dynamic (i.e., \emph{ramp} and \emph{duration}) and \emph{economic} flexibility analysis subject to the allocated network capacity to plan their market participation in more detail: this corresponds to determining the flexibility NOEs, which are in fact are a subset of capability and feasibility NOEs. The DERA \emph{technical} flexibility envelope calculated by each aggregator could then inform a distribution market operator (DMO) about their technical limitations, and the \emph{commercial} NOE could support a DERA in building its bid stack for multi-market participation. The DMO could then clear the local market and send the relevant dispatch signals to DERA. Besides, the DMO could sort the DERA bids and send the DERA pre-qualified bids to the upstream markets. Additionally, all these concepts could be used for DERA portfolio planning, e.g., to assess what resources an aggregator should acquire to provide a particular service.

	\vspace{-1em}
	\subsection{Implementation timescales and use cases}
	\vspace{-0.25em}
	The presented framework aims to conceptualise DER flexibility features and determine the relevant NOEs for a given time snapshot. Thus, all NOEs can in principle be calculated for each market clearance interval and for different market segments, e.g., day-ahead, real-time, etc. In this context, the NOE concept can be extended and implemented across different time scales from contingency analysis to real-time operation to operational planning and beyond to long term network planning. For example, a NOE-based contingency analysis could determine DERA network support potential~\cite{Dozein2019}, a real-time NOE could be useful for energy balance and FCAS allocation~\cite{Capitanescu2018b,Gonzalez2018,Heleno2015,Silva2018,Riaz19,Contreras2018}, and NOEs calculated for the next 24 hours could be used for dynamic allocation of network hosting capacity or for pre-dispatch analysis in day-ahead markets~\cite{Naughton2020}. Similarly, assessment from one week to a couple of years could be used for (short, medium and long term) system adequacy studies. Similarly, NOEs could also be incorporated in long-term planning and investment studies~\cite{Wang2019a}.
	
	\vspace{-1.5em}
	\subsection{Scalability and applications of the NOE approach}\label{Sec:Scalability NOE}\vspace{-0.25em}
	A key property of the NOE modelling approach is its \emph{flexibility} and \emph{scalability} across different network aggregation levels and could be deployed to estimate the potential upstream flexibility from the DER installed downstream. 
	
	\begin{figure}[]
		\centering
		\includegraphics[width=85mm] {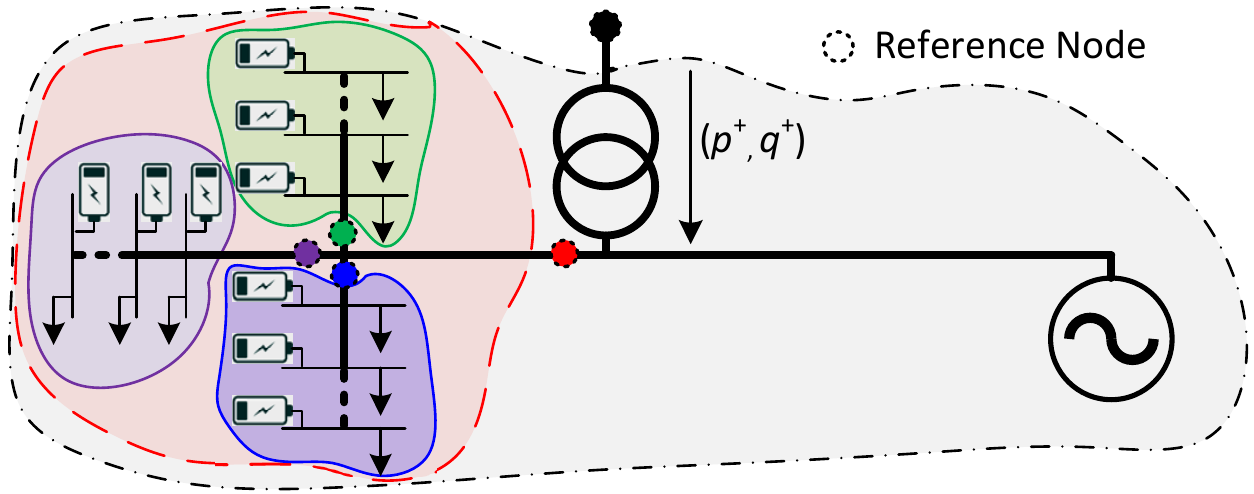}
		\vspace{-0.25em}
		\caption{Illustrative example of flexibility and scalability of NOEs.}
		\vspace{-1.50em}
		\label{fig:Exp_Net_Scalability}
	\end{figure}
	
	For example, the “steady-state” \emph{feasibility} NOE at the level of individual buildings or DER could be constructed from the demand and P-Q characteristics of individual devices to evaluate their aggregate power export/import limits. NOE at the level of an LV feeder (constructed from NOEs of individual buildings/DER) may be used to quantify the potential of the downstream DER for peak shaving or the time-varying hosting capacity of the feeder by Volt-VAr-Watt management. The \emph{hosting capacity} information of a feeder is in fact all implicitly contained in its feasibility NOE, which synthesised all the P-Q controllability of the downstream DER and loads subject to network constraints. In other words, the feasibility NOE can be used as a representation of the \emph{time-varying hosting capacity} of a feeder, network, etc., and is thus an operational extension (to be potentially deployed in real time or close to real time) of the concept of hosting capacity usually used for planning~\cite{Capitanescu2015,SeydaliSeyfAbad2020}. Similarly, the feasibility NOE from aggregation at the MV-LV transformer may be used to assess the reactive power support potential from the LV to the MV network. When moving up with voltage levels and in a similar fashion, relevant NOEs (which in turn may come from aggregation of downstream NOEs) may be used to assess the TSO-DSO flexibility potential\footnote{Furthermore, it may be applied to aggregation levels with multiple reference points. However, for simplicity we will refer to only one grid supply point, as typical in distribution networks that are operated radially.}. For these applications, however, it is likely that other types of NOE with representation of \emph{dynamic} flexibility features too may be more relevant, for example the \emph{ramp} or \emph{duration} envelopes for provision of different services. Finally, the defined \emph{economic}, \emph{technical} and \emph{commercial} NOEs can be deployed by aggregators and DSO within a distributed marketplace and for participation in upstream markets.

\begin{figure}[]
		\centering	
		\includegraphics[width=\linewidth] {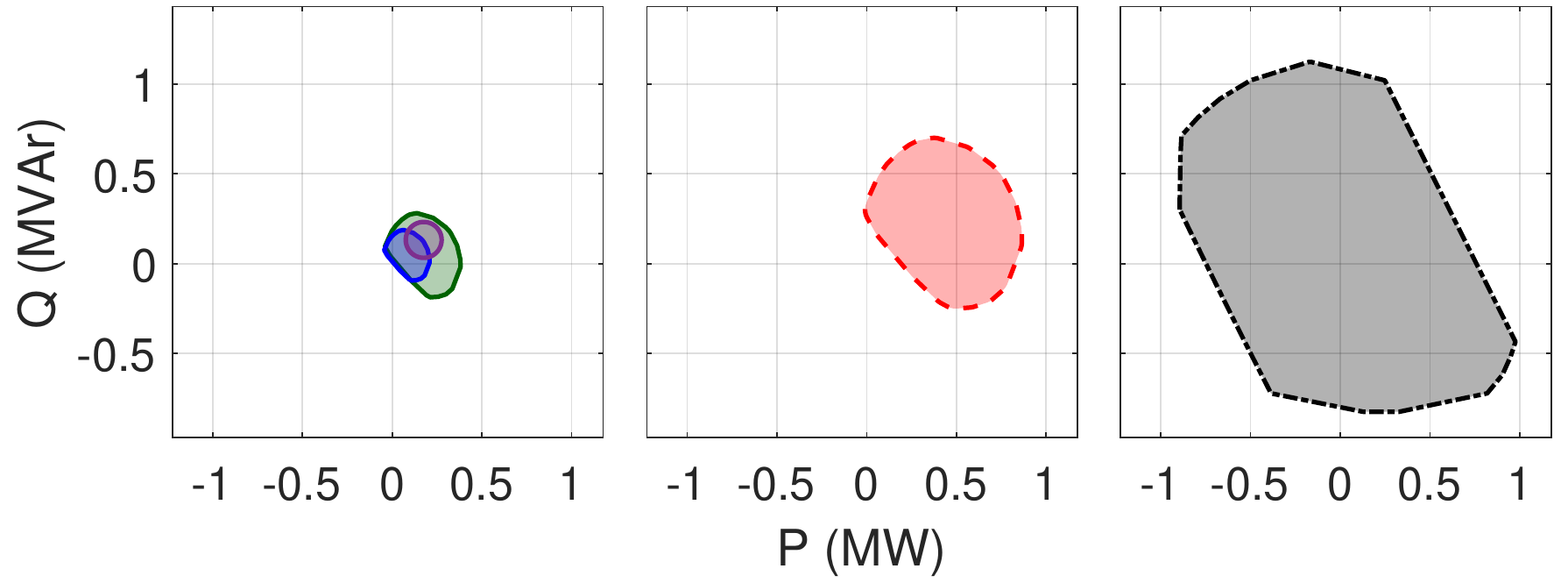}
		\vspace{-2.0em}
		\caption{Feasibility NOEs for the three individual feeders (left), aggregation of the three feeders (middle), and entire system (right) shown in Fig.~\ref{fig:Exp_Net_Scalability}.}
		\vspace{-0.75em}
		\label{fig:Scalability_NOE}
	\end{figure}

	For example, Fig.~\ref{fig:Exp_Net_Scalability} shows a modified version of the DERA described in Section~\ref{Sec:Exp}, where the load and BESS are spread across the green, blue and purple feeders and aggregated first at each feeder level, with their respective green, blue and purple reference nodes shown in Fig.~\ref{fig:Exp_Net_Scalability}. The NOEs associated with the three feeders are shown in Fig.~\ref{fig:Scalability_NOE} (left). The NOE at the red-reference node in Fig~\ref{fig:Exp_Net_Scalability} represents the aggregated capacity of all three feeders, shown in Fig.~\ref{fig:Scalability_NOE} (middle), which is then aggregated with the generator to establish the feasibility NOE of the entire system (computed at black-reference node mentioned in Fig.~\ref{fig:Exp_Net_Scalability}), shown in Fig.~\ref{fig:Scalability_NOE} (right). This illustrates the scalability of the NOE approach introduced, whereby reference nodes can be flexibly defined and moved across different aggregation levels.	
	
	\vspace{-.750em}	
	\section{Nodal Operating Envelope Modelling Framework}\label{Sec:NOE framework}
\begin{figure}[]
		\centering
		\includegraphics[width=\linewidth] {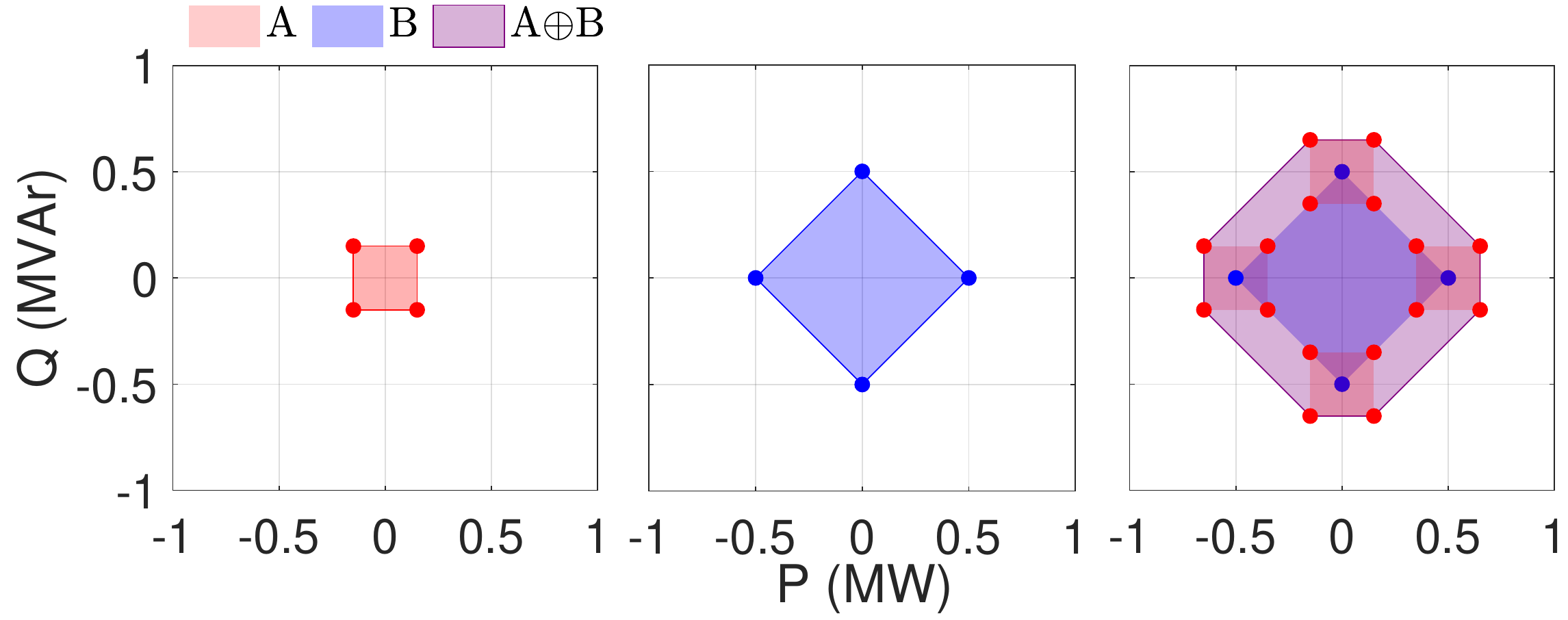}
		\vspace{-2.25em}
		\caption{Minkowski addition of two sets in P-Q space.}
		\vspace{-1.75em}
		\label{fig:Mink_Sum}
	\end{figure}

	%\vspace{-0.25em}
	The NOEs of a \emph{virtual} DERA can be readily obtained by Minkowski summation of the individual DER modelled via set representation~\cite{Chicco2020,Hinker2018}. For example, for any two sets represented by their position vectors $\boldsymbol{a}$ and $\boldsymbol{b}$, the Minkowski Sum (denoted by $\oplus$) is given by~\eqref{MinkSum}, and a graphical example for P-Q space is shown in~Fig.~\ref{fig:Mink_Sum}~\cite{Wein2016}.
	\vspace{-0.5em}
	\begin{equation} \label{MinkSum}
		A\oplus B = {\boldsymbol{a} + \boldsymbol{b} : \boldsymbol{a} \in A, \boldsymbol{b} \in B}.
		\vspace{-0.5em}
		\end{equation}
		However, for a \emph{physical} DERA, the calculation of operating envelopes is much more complex as the network introduces nonlinear coupling constraints between DER and loads.
	\vspace{-1.25em}
	\subsection{General modelling options considering network} \vspace{-0.25em}
	Previous methodologies to estimate aggregated flexibility from downstream DERs while considering the presence of the network only addresses \emph{feasibility} NOE (i.e., P-Q chart) and only in the context of TSO-DSO. These methodologies may be broadly based on either optimisation approaches~\cite{Capitanescu2018b,Silva2018,Contreras2018} or Monte Carlo based simulations~\cite{Heleno2015,Wang2019a,Gonzalez2018,Riaz19} and can be modified to calculate other types of NOEs, as introduced in Section~\ref{Sec:FLC_xtics}.
	
	\emph{Optimisation} approaches explicitly consider DER and network constraints and deploy AC-OPF analysis to obtain a single point on the NOE boundary. First, four AC-OPF runs are executed with objectives to find the minimum and maximum active and reactive power import limits. This provides the minimum bounding box around a NOE. Then, the NOE can be determined either by exploring boundary point along various directions around the operating point~\cite{Silva2018} or by dividing the range of reactive power in equal intervals and solving for the active power range (i.e., minimum and maximum) within each interval~\cite{Capitanescu2018b}. The resultant order set of active power and reactive power then define the NOE.
	
	\emph{Monte Carlo-based} simulations rely on a large number of power flow instances formed by randomly generated DER operational points. The points resulting in violation of network constraints are discarded, and the convex hull of remaining points provides the NOE. The Monte Carlo approach is simple and can generally be applied to various topologies. However, it is inherently time demanding for accurate estimation of NOE, since the number of power flow instances required grows exponentially with the number of DER and parameter of interest while only a few of the randomly generated points lie on the envelope boundary itself.
	
	The presented methodology extends the optimisation approach~\cite{Capitanescu2018b,Silva2018} and builds on~\cite{Wang2019a,Riaz19} to effectively represent different \emph{flexibility} NOEs, and advances the state of the art by further developing the optimisation-based approach to build NOEs for feasibility, ramp, duration, economic, technical and commercial flexibility features in a single coherent framework, outlining the role of each feature as summarised through the Venn diagram in Fig.~\ref{fig:Flx_VenDia}.
	
	\vspace{-1.00em}
	\subsection{Mathematical model}\label{Sec:Math Model}
	\vspace{-.25em}	
	\subsubsection{Objective function}
	The framework employs one of the four following objective function~\eqref{Eq:Pmin}-\eqref{Eq:Qmax}, aiming to minimise/maximise the active ($p^{\text{+}}$) and reactive ($q^{\text{+}}$) power import of DERA, depending upon the requirement.
	%\vspace{-0.25em}
	\begin{align}
	&\mathop{\operatorname{minimise}} \limits_{p_r,q_r,\Delta p_r,\Delta q_r,\zeta_d, t_{ij},v_i, \theta_i} \quad & p^{\text{+}}\label{Eq:Pmin} \text{,}\\
	&\mathop{\operatorname{maximise}} \limits_{p_r,q_r,\Delta p_r,\Delta q_r,\zeta_d, t_{ij},v_i, \theta_i} \quad & p^{\text{+}}\label{Eq:Pmax}\text{,}\\
	&\mathop{\operatorname{minimise}} \limits_{p_r,q_r,\Delta p_r,\Delta q_r,\zeta_d, t_{ij},v_i, \theta_i} \quad & q^{\text{+}}\label{Eq:Qmin}\text{,}\\
	&\mathop{\operatorname{maximise}} \limits_{p_r,q_r,\Delta p_r,\Delta q_r,\zeta_d, t_{ij},v_i, \theta_i} \quad & q^{\text{+}}\label{Eq:Qmax}\text{,}
	%\vspace{-0.25em}
	\end{align}
	where DER active/reactive power dispatch ($p_r$/$q_r$) and deviation ($\Delta p_r$/$\Delta q_r$), demand curtailment factor ($\zeta_d$), on load tap changer (OLTC) transformer ratio ($t_{ij}$), voltage magnitude ($v_i$) and voltage angle ($\theta_i$) of each bus are the decision variables. On the other hand, response time ($\tau$), call length ($\psi$), flexibility cost DERA willing to pay ($c$), dispatch point ($[p_r^\lambda,q_r^\lambda]$), minimum/maximum active ($\overline{p}_r/\underline{p}_r$) and reactive power limits ($\overline{q}_r/\underline{q}_r$), activation/deactivation time ($t_r^\text{act-/+}$), ramp up/down rate $r_r^\text{-/+}$, energy limits ($\overline{e}_r/\underline{e}_r$), state of charge ($e_r$), and active/reactive power deviation cost ($\rho_r^p/\rho_r^q$)  of each resource serve as the input parameters to the optimisation framework, for a (parametric) calculation of the relevant NOE. Moreover, the choice of parameters response time ($\tau$), call length ($\psi$) and maximum cost DERA is willing to pay ($c$) depends upon the required granularity of the corresponding flexibility NOE.

	\subsubsection{Constraints}
	The above optimisation problem is constrained by network restrictions and DER operational limitations. The model deploys general AC power flow constraints along with the line thermal, bus voltage, and branch angle difference limits (e.g., as presented in~\cite{Capitanescu2018b}). The active and reactive loads are modelled according to voltage-dependent exponential loads~\cite{Milanovic1999}. Furthermore, all DER are constrained by maximum and minimum active/reactive power (e.g., synchronous generator’s capability map, BESS maximum charge and discharge rates, etc.), power factor, and inverter rating. 
	
	DER aggregated power deviation can be expressed as:%~\eqref{Eq:DelPr} and~\eqref{Eq:DelQr}:
	\vspace{-0.25em}  
	\begin{equation}\label{Eq:DelPr}
	\Delta p_r = p_r - {p}_r^\lambda \quad \forall r \in \mathcal{R} \text{,}
	\vspace{-0.25em} 
	\end{equation}	
	\begin{equation}\label{Eq:DelQr}
	\Delta q_r = q_r - {q}_r^\lambda \quad \forall r \in \mathcal{R} \text{,} 
	\vspace{-0.25em}
	\end{equation}
	where $\Delta p_r$/$\Delta q_r$ represents the active/reactive power deviation, and ${p}_r^\lambda$/${q}_r^\lambda$ is the active/reactive power dispatch of the resource $r$.
	Active power deviation is limited by ramp-up (${r}_r^{\text{+}}$) and down rates (${r}_r^{\text{-}}$):%~\eqref{Eq:Ramp}:
	\vspace{-0.25em}
	\begin{multline}\label{Eq:Ramp}
	\min(0,- (\tau-{t}_r^{\text{act-}}) {r}_r^{\text{-}}) \leq \Delta p_r \leq \\\max(0, (\tau-{t}_r^{\text{act+}}) {r}_r^{\text{+}}) \quad \forall r \in \mathcal{R} \text{,}
	\vspace{-0.25em} %\frac{\overline{P}}{t_r^{\text{r}+}}
	\end{multline}
	where the ${t}_r^{\text{act+/-}}$ represents the time to activate/deactivate the resources and includes delaying factors such as communication delay, start-up time, etc. The $\min$/$\max$ conditions restrict the corresponding lower and/or upper bounds of $\Delta p_r$ to zero in case $\tau\leq{t}_r^{\text{act-/+}}$.

	The active power deviation of storage units is also restricted by the amount of energy that is required to fulfil an energy market commitment (${p}_r^{\lambda} \Delta t$) and sustain the deviated operating point ($\Delta p_r \psi$) for the maximum service duration $\psi$:
	\vspace{-0.25em}
	\begin{equation}\label{Eq:Enrg}
	\underline{{e}}_r \leq {e}_r - {p}_r^\lambda \Delta t - \Delta p_r \psi \leq \overline{{e}}_r  \quad \forall r \in \mathcal{B} \text{,}
	\vspace{-0.25em}
	\end{equation}
	where $\mathcal{B}$ is the set of storage units ($\mathcal{B} \subseteq \mathcal{R}$), ${e}_r$ is the state of charge, and $\Delta t$ is the energy market dispatch interval. Constraint~\eqref{Eq:Enrg} thus ensures that storage minimum/maximum state of charge limits ($\underline{e}_r$/$\overline{e}_r$) are not exceeded during service provision. 
	The cost of deviation for each resource can be captured through~\eqref{Eq:DERCost} as proposed by~\cite{Zhang2005,Silva2018}.

	\vspace{-0.25em}
	\begin{equation}\label{Eq:DERCost}
	\sum_{r \in \mathcal{R}}  \rho_r^\text{p} |\Delta p_r| + \rho_r^\text{q} |\Delta q_r| \leq c \text{,}
	\vspace{-.5em}
	\end{equation}
	where $\rho_r^\text{p/q}$ is the cost of active/reactive power deviation per \SI{}{\mega\watt\hour}/\SI{}{\mega\var\hour}, and $c$ is the (maximum) cost associated with deployment of a certain amount of flexibility.
	
	\vspace{-1.25em}
	\subsection{Methodology to build NOEs for various flexibility features} \label{Sec:RSE_Metodology}
		\vspace{-0.25em}
		The methodology to estimate NOEs, including network constraints, for feasibility, ramp, duration, economic, technical and commercial flexibility generates $4+2K$ P-Q boundary points for a given combination of response time ($\tau$), call length ($\psi$) and cost ($c$). Multiple runs of the methodology with different combinations of parameter $\tau$, $\psi$ and $c$ can then approximate the relevant NOE for a given time interval. The value of $K$ is selected to strike a balance between accuracy and computational complexity (later discussed in the case study), whereas the parameters $\tau$, $\psi$ and $c$ can be selected based on market specifications. Each boundary point is obtained by solving the optimisation problem presented in Section~\ref{Sec:Math Model} considering, relevant parameters for each specific NOE of interest, the AC-OPF constraints (e.g., see~\cite{Capitanescu2018b}) along with relevant constraints presented in Table~\ref{Table:constraints}.
		\begin{table}[]
			% increase table row spacing, adjust to taste
			\renewcommand{\arraystretch}{1.3}
			\caption{Relevant constraints for specific NOE evaluation.}
			\vspace{-.5em}
			\label{Table:constraints}
			\centering
			% Some packages, such as MDW tools, offer better commands for making tables
			% than the plain LaTeX2e tabular which is used here.
			\begin{tabular}{|l|c|}
				\hline
				Nodal operating envelope & Constraints\\
				\hline
				Feasibility & \eqref{Eq:DelPr}, \eqref{Eq:DelQr}\\
				\hline
				Ramp flexibility & \eqref{Eq:DelPr}-\eqref{Eq:Ramp}\\
				\hline
				Duration flexibility & \eqref{Eq:DelPr},~\eqref{Eq:DelQr},~\eqref{Eq:Enrg}\\
				\hline
				Economic flexibility & \eqref{Eq:DelPr},~\eqref{Eq:DelQr},~\eqref{Eq:DERCost}\\
				\hline	
				Technical  flexibility\footnote{} & \eqref{Eq:DelPr}-\eqref{Eq:Enrg}\\
				\hline
				Commercial flexibility\footnote[3]{} & \eqref{Eq:DelPr}-\eqref{Eq:DERCost}\\
				\hline		
			\end{tabular}	
			\vspace{-1.0em}
		\end{table}
	\footnotetext[3]{Market-specific constraints other than response time and call length can readily be added as per market/service requirements.}
	
	The main Steps to compute one P-Q slice of NOE are:
		\begin{enumerate}[leftmargin=*]
			\item Calculate the four extreme boundary points of the relevant P-Q slice by solving four optimisation routines considering one objective function at a time among~\eqref{Eq:Pmin}-\eqref{Eq:Qmax}.\label{RSE:Step1}	
			\item 	Using the values of maximum and minimum reactive powers ($\overline{q}^{\text{+}}$/$\underline{q}^{\text{+}}$) obtained in Step~\ref{RSE:Step1}, compute $K$ equidistant intermediate reactive power points using~\eqref{Eq:Qk}. The value of $K$ can be selected based on the required granularity of the NOE. For example, to obtain NOE with $N$ point per \SI{}{\mega \var}, $K$ can be calculated as $K=N(\overline{q}^+ - \underline{q}^+)$. \label{RSE:Step2}
			%\vspace{-.25em}
			\begin{equation}\label{Eq:Qk}
			q_k^\text{+}=\underline{q}^\text{+} + k(\overline{q}^\text{+}-\underline{q}^\text{+})/K \quad \forall k=1\dots K \text{.}
			%\vspace{-.25em}
			\end{equation}	
			\item For each value of $k=1\dots K$, solve an optimisation problem with objective function~\eqref{Eq:Pmin} subject to the relevant constraints of Table~\ref{Table:constraints} and the additional constraint~\eqref{Eq:QBinding}. The role of this constraints is to fix the reactive power import at $q_k^{\text{+}}$, thus providing intermediate boundary points between maximum and minimum values of reactive power, while minimising the active power import of DERA.\label{RSE:Step3}
			%\vspace{-0.25em}
			\begin{equation}\label{Eq:QBinding}
			q^\text{+} = q_k^\text{+} \text{.}
			%\vspace{-0.25em}
			\end{equation} 	
			\item Repeat Step~\ref{RSE:Step3} using objective function~\eqref{Eq:Pmax}, to find the maximum active power import of DERA for the corresponding value of reactive power.\label{RSE:Step4}
	\end{enumerate}
		
	Note that the methodology generates NOE estimates for a single operating snapshot (e.g., dispatch interval). Notably, to speed up the assessment the optimisation runs of Step~\ref{RSE:Step1} can be performed in a parallel fashion, and so can the runs in Step~\ref{RSE:Step3} and~\ref{RSE:Step4}.

	\vspace{-1.0em}
	\section{Case Study Applications}\label{Sec:Case Study}
	\subsection{Test Network}
\begin{figure}[]
		\centering
		\includegraphics[width=\linewidth] {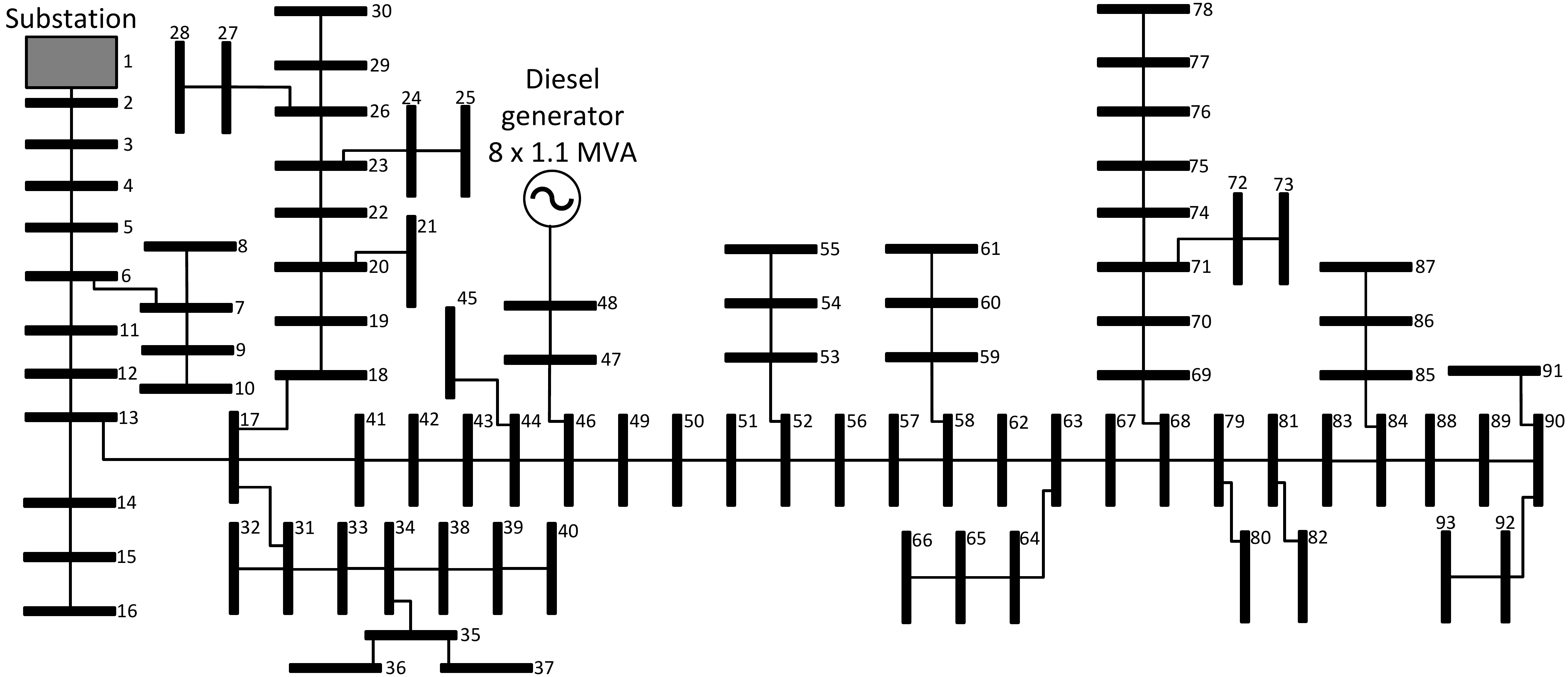}
		\vspace{-1.5em}
		\caption{Distribution network used for the case study.}
		\vspace{-1.5em}
		\label{fig:CS_Network}
	\end{figure}

	\vspace{-0.25em}
	The efficacy of the proposed methodology is demonstrated on a real Australian MV distribution network, as shown in Fig.~\ref{fig:CS_Network}. The network consists of 93 nodes, operates at~\SI{22}{\kilo\volt}, and serves a total of 3136 customers out of which 695 (~\SI{22}{\percent}) have rooftop PV with a combined capacity of ~\SI{2.3}{\mega\voltampere}. Furthermore, the customers on average have installed~\SI{1}{\mega\watt\hour} of storage against each~\SI{}{\mega\watt} of installed PV. The capacity of the distribution transformers is~\SI{12.9}{\mega\voltampere} Furthermore, eight identical diesel generation units of~\SI{1.1}{\mega\voltampere} each with an activation delay of~\SI{25}{\second} and ramp rate of ~\SI{18.3}{\kilo \watt \per \second} are connected in the network. It is assumed that for the time interval under consideration, all diesel generators are offline and the average state of charge of batteries is~\SI{30}{\percent}. Furthermore, the net power demand of the loads is~\SI{7.4}{\mega\watt} and~\SI{3.1}{\mega\var} with negligible generation from rooftop PV. The nominal voltage bound is assumed $\pm$~\SI{5}{\percent}.

	In general, different DER might be controlled by different and multiple aggregators. Moreover, only a portion of the network might be considered in the analysis, for example where a specific distributed marketplace has been developed at a community level. However, as the case study aims at a fundamental demonstration of the proposed methodology to assess different NOE and flexibility features, for simplicity we consider here that the entirety of the distribution network is involved in the analysis, that is, all DER are assumed to be controlled by one aggregator and interacting with one DSO.
	
	The methodology has been implemented using AMPL algebraic language and solved using the nonlinear solver Knitro in a parallel fashion on a system with i7-7820HQ CPU (with four physical cores) @ 2.90 GHz and 32 GB RAM. 
	
	\vspace{-1.25em}
	\subsection{Market/support services description}
	\vspace{-0.25em}
	Three different markets/network support services are considered in the analyses below:
	
	\subsubsection{Symmetric reserve}
	some markets do not differentiate between upward and downward reserves (e.g., secondary reserve in Europe), and only allow symmetric bids. Therefore, in symmetric reserve markets, all participants are required to provide equal capacity for upward and downward flexibility.
	
	\subsubsection{DR}
	all participants are required to reduce their power import from the grid for a specific duration. For such services there is usually a certain notice time and the service duration could be several hours. In the analysis, we consider short-DR with duration of~\SI{30}{\minute} and long-DR with duration of ~\SI{4}{\hour}.
	
	\subsubsection{Contingency frequency controlled ancillary services (FCAS)}
	In the Australian National Electricity Market (NEM) there are six contingency FCAS markets, i.e., fast raise/lower, slow raise/lower, and delayed raise/lower. The participants in fast, slow, and delayed FCAS are required to respond within~\SI{6}{\second},~\SI{60}{\second}, and~\SI{300}{\second} of a contingency and sustain this response for~\SI{60}{\second},~\SI{5}{\minute}, and~\SI{10}{\minute}, respectively.
	
	\vspace{-1.0em}
	\subsection{Scalability and accuracy of the approach}\label{Sec:Results}
\begin{figure}[]
			\centering
			\includegraphics[width=70mm] {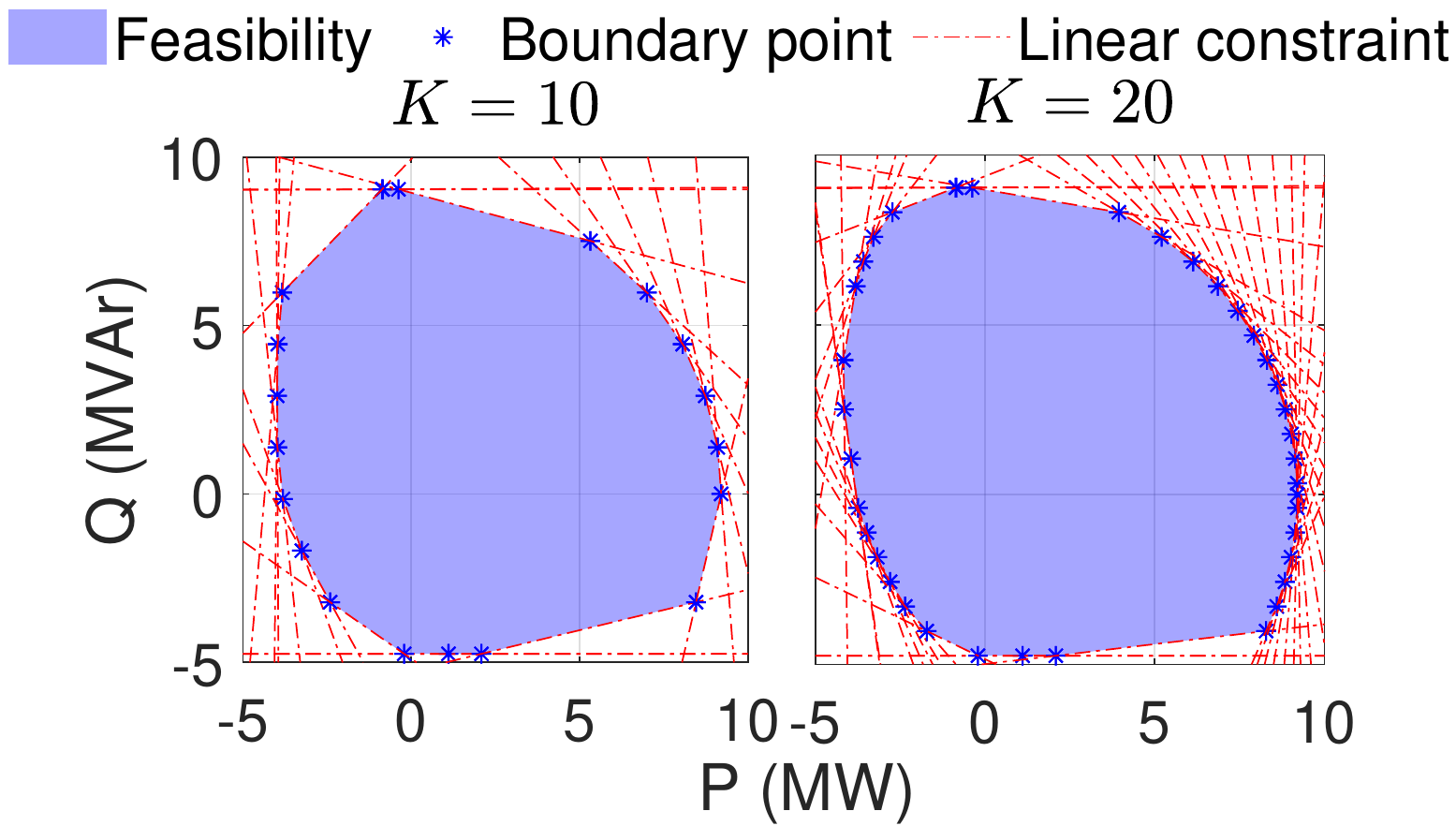}
			\vspace{-0.75em}
			\caption{Boundary points, corresponding linear constraints, and resulting feasibility NOE, with K=10 (left) and K=20 (right) boundary points.}
			\vspace{-0.5em}
			\label{fig:CS_Linear_constraints}
		\end{figure}

\begin{figure}[]
			\centering
			\includegraphics[width=\linewidth] {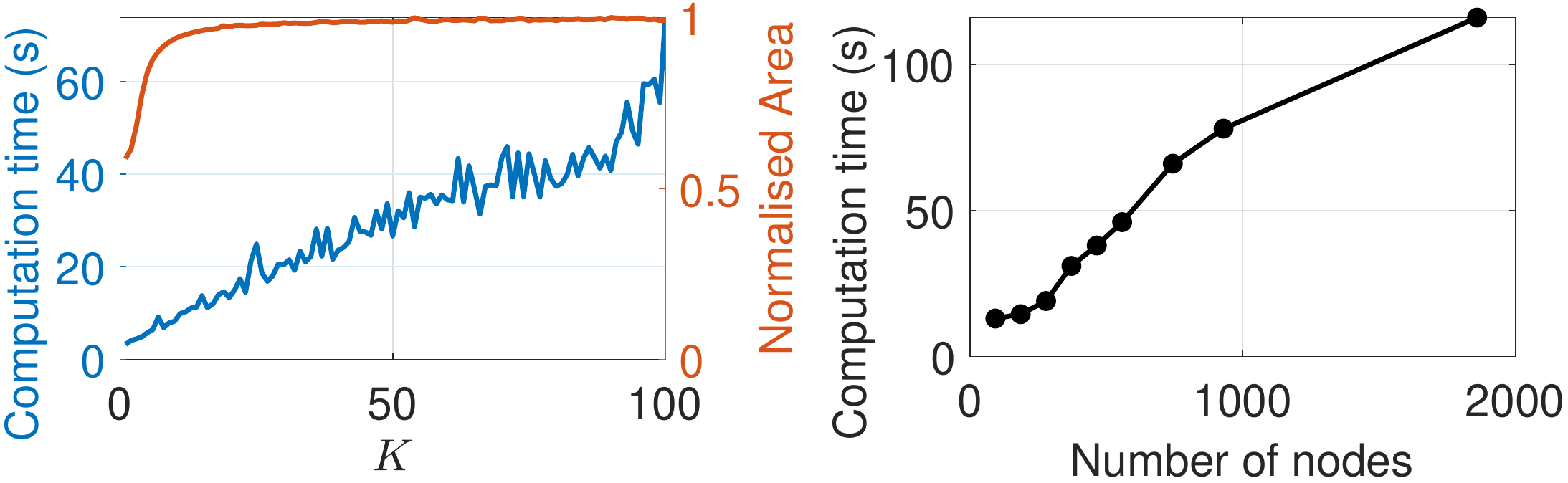}
			\vspace{-1.75em}
			\caption{Impact of $K$ on the NOE accuracy and computation time of test network (left), and computational time as a function of the number of system nodes (right) for $K=20$.}
			\vspace{-1.5em}
			\label{fig:CS_Impact_K}
			%\vspace{-0.5em}
		\end{figure}

		\vspace{-0.25em}
		The methodology presented in Section~\ref{Sec:RSE_Metodology} generate $2K+4$ boundary points and the consecutive boundary points can then be deployed to generate linear binding constraints for different flexibility envelopes. Therefore, the NOEs can be defined mathematically through a set of linear constraints of the form, $\boldsymbol{A}\boldsymbol{x}\leq\boldsymbol{b}$, which can be easily incorporated in the calculation of NOEs at some upstream node. The presented methodology is thus \emph{conceptually} scalable to calculate NOEs at different aggregation levels, as discussed in Section~\ref{Sec:Scalability NOE}. In addition, these NOEs (as a set of linear constraints) at appropriate aggregation levels can also be included by the system operator in the market clearing engine. The feasibility boundary points along with the corresponding linear binding constraints for our DERA example are shown in Fig.~\ref{fig:CS_Linear_constraints} for $K=10$ (left) and $K=20$ (right).
		As mentioned earlier, the NOE area can be refined by selecting an appropriate value of $K$ to strike a balance between accuracy and computation complexity. In this respect, the results in Fig.~\ref{fig:CS_Impact_K} (left)
		represent the computation time and the normalised feasibility envelope area (the area for $K=100$ is used as a benchmark normalisation factor), as a function of $K$. The NOE area initially exponentially increases and then saturates. This suggests that, for this specific case, the NOE may not be accurately captured with $K\leq20$, while higher values (i.e., $K\geq40$) would not necessarily improve the accuracy but result into higher computation time. In general, such tradeoff studies would need to be run case by case depending on the network and DER portfolios under consideration, and a few sensitivities could be run very quickly by spanning across several values of $K$, for example at increments of $10$).
	
	Another point of interest in terms of computational scalability is the parallelisation potential of the approach, as for significantly large networks the NOE can be computed by first calculating the NOEs of appropriate sub-networks and then the NOE of the entire system from the NOEs of these sub-networks. For example, if we increase the size of the case study by replicating the test network feeder (Fig.~\ref{fig:CS_Network}), the resulting computation time is reported in Fig.~\ref{fig:CS_Impact_K} (right). The results indicate that for a system with $20$ such feeders ($1860$ node), the total computation time is \SI{116}{\second}. However, by deploying distributed computing, the NOE of each feeder could be simultaneously calculated within \SI{13}{\second} (see Fig.~\ref{fig:CS_Impact_K} (right) for 93 nodes) and then the NOE at the substation could be calculated in \SI{10}{\second}. This would thus overall reduce the computation time from \SI{116}{\second} to some \SI{23}{\second}. However, in general, each distribution system is unique in terms of network structure, number and location of DER, etc.; hence, parallelisation benefits would need to be assessed case by case.
	\vspace{-2.5em}
	\subsection{Flexibility features and NOEs discussion}
\begin{figure}[]
		\centering
		\includegraphics[width=85mm] {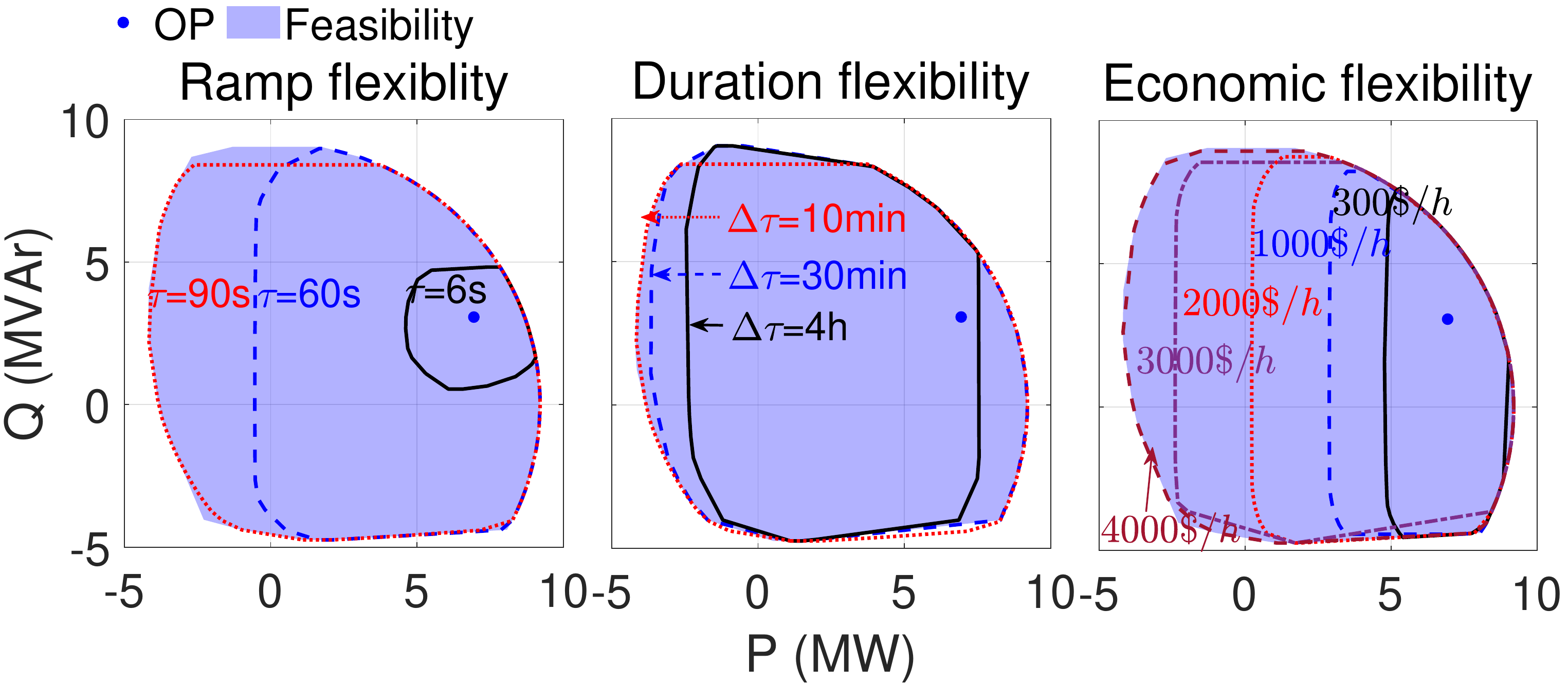}
		\vspace{-1em}
		\caption{Ramp (left), duration (middle) and economic (right) flexibility NOEs.}
		\vspace{-1em}
		\label{fig:CS_RSEFlx}
	\end{figure}

	\vspace{-0.25em}
	The general ramp (left), duration (middle) and economic (right) flexibility NOEs for the condition under study are given in Fig.~\ref{fig:CS_RSEFlx},
	for different values of response time ($\tau$), call length ($\psi$) and maximum cost ($c$), along with the operational point (OP), and feasibility NOE.
	
\begin{figure}[]
		\centering
		\includegraphics[width=\linewidth] {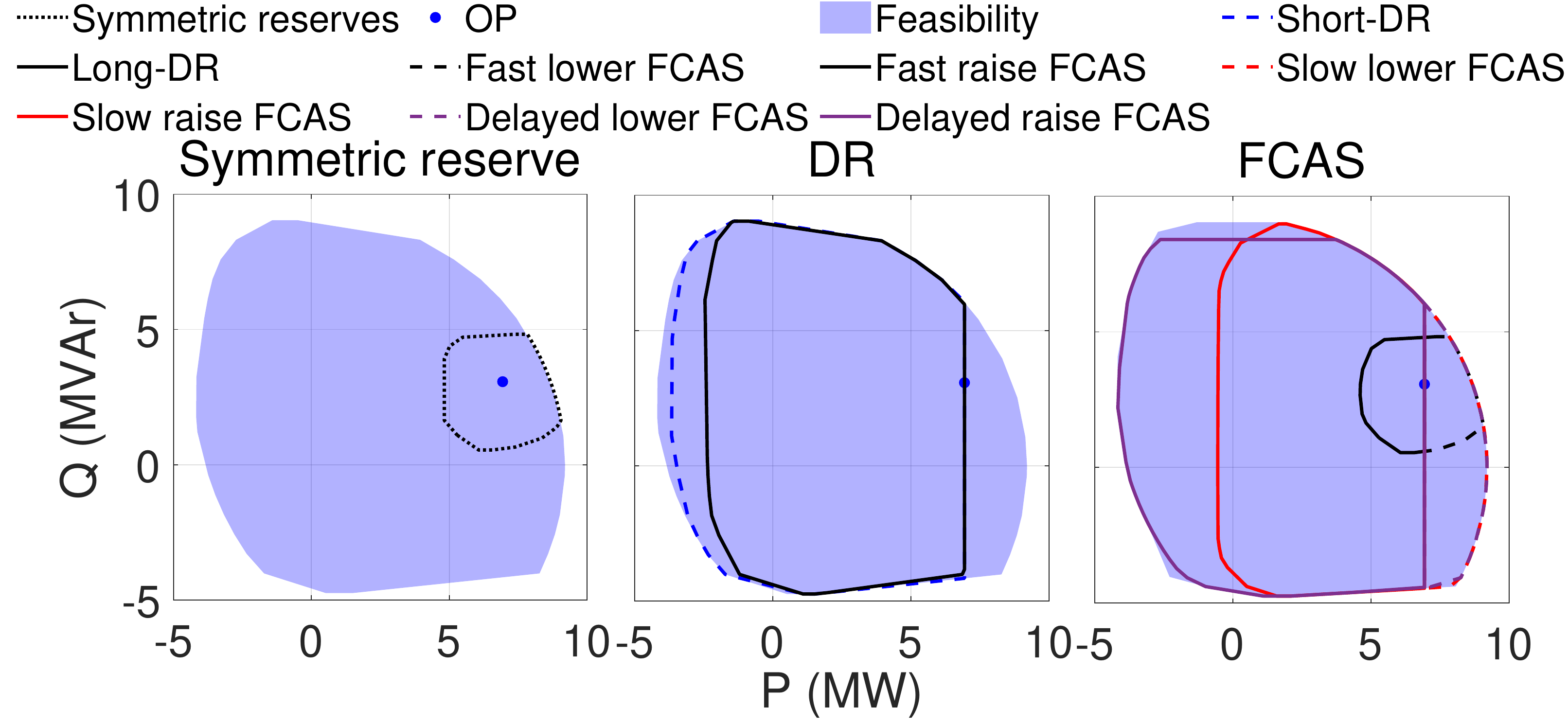}
		\vspace{-2em}
		\caption{Technical flexibility NOEs for symmetric (left), DR (middle) and FCAS (right) services.}
		\vspace{-1.75em}
		\label{fig:CS_TFlx}
	\end{figure}

	Although these flexibility features provide valuable information to an aggregator for both operation and planning purposes, it is the \emph{technical} flexibility that defines the DERA technical ability to participate in a particular market segment,and therefore serves as the necessary and sufficient set of linear constraint for the market operator to accurately account for DERA participation potential. The \emph{technical flexibility} NOE for symmetric reserve, DR and FCAS are shown in Fig.~\ref{fig:CS_TFlx},
	along with DERA feasibility NOE. In this particular case, the DERA has more potential to provide upward than downward flexibility, as all the diesel generators are offline. However, due to the symmetry requirement, it is only able to offer a limited flexibility for the symmetric reserve, as shown in Fig.~\ref{fig:CS_TFlx} (left). In contrast, the DR service aims to reduce the power import from the grid (i.e., upward flexibility), and the DERA can provide its maximum potential, restricted only by the energy content of the resources (i.e., duration flexibility). The technical flexibility envelopes illustrate a higher potential to provide short-DR than long-DR, as shown in Fig.~\ref{fig:CS_TFlx} (middle), as the operating point deviation for batteries is more restricted for long-DR. Contingency FCAS allow asymmetric bids, and so the DERA can offer the optimal volume for each market segment, as shown in Fig.~\ref{fig:CS_TFlx} (right). In case of raise FCAS, DERA participation is mostly restricted by the ramp flexibility, as diesel generators need time to start and ramp-up, while only batteries can provide lower FCAS given the fast response time requirement of~\SI{6}{\second}.
	
\begin{figure}[]
		\centering
		\includegraphics[width=75mm] {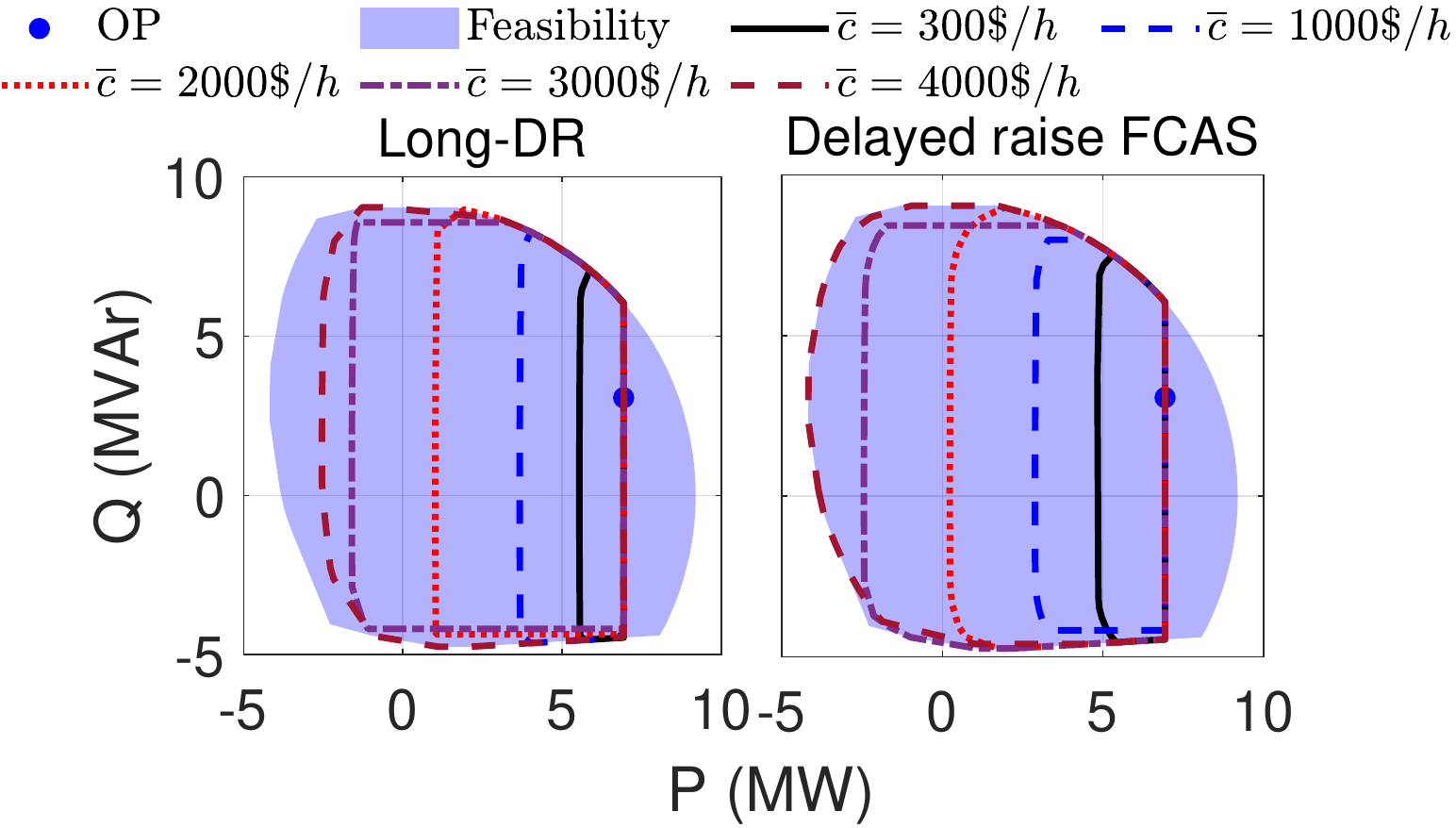}
		\vspace{-1.0em}
		\caption{Commercial flexibility NOEs for long-DR (left) and delayed raise FCAS (right).}
		\vspace{-1.75em}
		\label{fig:CS_CFlx}
	\end{figure}

	While the technical envelopes provide sufficient information to demonstrate the technical ability of DERA to participate in a given market, the commercial envelopes allow a complete techno-economic cost-benefit analysis. As example, the \emph{commercial flexibility} NOEs of the long-DR (left) and delayed raise FCAS (right) services are shown in Fig.~\ref{fig:CS_CFlx}.
	Note that the commercial flexibility envelopes of delayed FCAS closely resembles the economic flexibility envelopes, whereas for the long-DR this is not the case. In this particular case, for the delayed FCAS, ramp and duration flexibility are in fact not acting as binding constraints, because all the DER can ramp up to the maximum output levels within~\SI{300}{\second}/\SI{5}{\minute} and have sufficient energy to sustain the maximum deviation for~\SI{600}{\second}/\SI{10}{\minute}. Therefore, in calculating the commercial flexibility for delayed raise FCAS, the active constraints are identical to the economic flexibility ones, resulting in identical envelopes. However, for the case of long-DR the batteries, which are also the cheaper resource, do not have adequate energy to sustain maximum power output for the call length of~\SI{4}{\hour}. The commercial flexibility envelope of long-DR is thus more constrained, resulting into a higher operational cost, as shown in Fig.~\ref{fig:CS_CFlx}. In summary, while the economic flexibility NOEs indicate the minimum cost of operating point deviation of a DERA, it is the commercial flexibility NOEs to provide the actual cost to assess DERA variability in a given market.
	
\begin{figure}[]
		\centering
		\includegraphics[width=58mm] {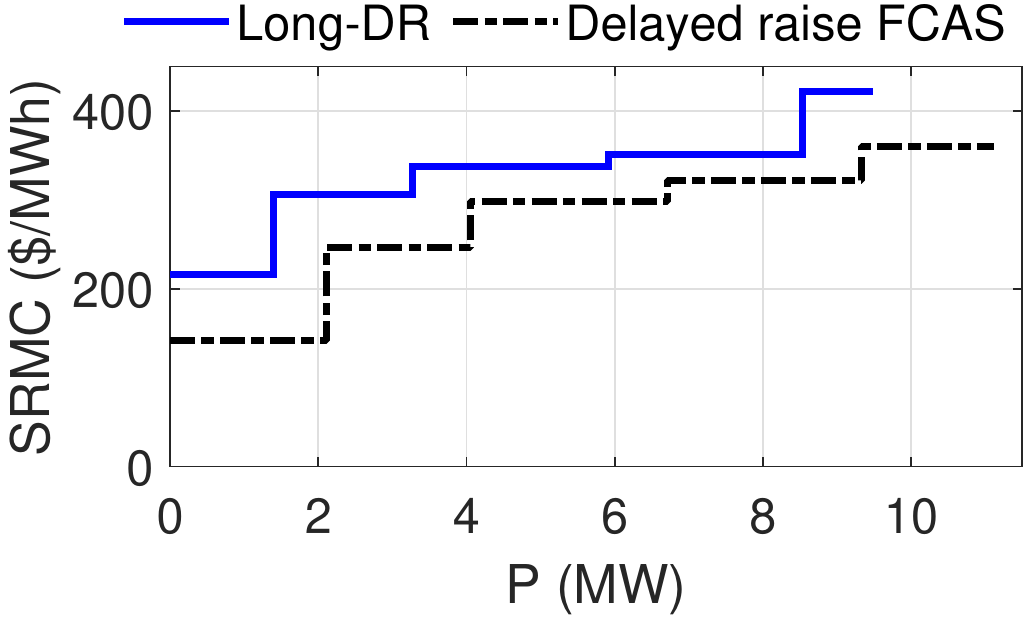}
		\vspace{-1em}
		\caption{Bid stack of the test DERA for long-DR and delayed raise FCAS.}
		\vspace{-1.5em}
		\label{fig:CS_SRMC}
	\end{figure}

	Furthermore, the short run marginal cost (SRMC) of a DERA can be calculated as the ratio of cost ($c$) and active power range of each commercial flexibility envelope. In a competitive market arrangement, the SRMC also represents the bid price (y-axis) against the bid volume (x-axis) of a DERA. Therefore commercial flexibility NOEs aid in the creation of a DERA bid stack for an individual market segment. The bid stack of the DERA for long-DR and delayed raise FCAS services is shown in Fig.~\ref{fig:CS_SRMC}.
	Note that the presented methodology is not intended to optimise the participation of a DERA into multiple markets, but rather aims to understand the interaction and possible uses cases and features of various fundamental aspects of DER flexibility. 
	
	In conclusion, the information provided by different NOE holds the key to evaluate DERA market potential, design resource portfolio, and establish sustainable business cases. For example, a comparison of Fig.~\ref{fig:CS_RSEFlx} and Fig~\ref{fig:CS_TFlx} reflect that the DERA FCAS participation is limited by the ramp flexibility while its DR participation is restricted by the duration flexibility. Similarly, the comparison between the economic and commercial flexibility (Fig.~\ref{fig:CS_RSEFlx} and Fig.~\ref{fig:CS_CFlx}) highlights the increased flexibility provision cost when subject to a particular market requirement. For example, in the case when a cheaper unit is restricted due to market requirement and instead an expensive unit is dispatched to deliver the service. Such a comparison allows DERA to evaluate and priorities its participation in multiple services from a cost viewpoint.
	
	\vspace{-0.50em}
	\section{Conclusion}\label{Sec:Conclusion}
		%\vspace{-0.25em}
		This paper has introduced a foundational modelling and assessment framework, demonstrated and illustrated by the concept of nodal operating envelope (NOE), to assess various DER flexibility features while also accounting for reactive power and network constraints. Starting from differentiating between virtual and physical aggregations of DER, the concepts of capability, feasibility, ramp, duration, economic, technical, and commercial flexibility have been introduced and represented mathematically through a set of constraints emerging from a Venn diagram and represented visually through P-Q operational maps (a powerful tool to represent the relevant flexibility NOEs). Specific methodologies based on OPF analysis have also been presented to build each NOE. These concepts may be relevant to different use cases and stakeholders (e.g., aggregators, distribution system operators, distribution market operators, etc.), and are key to assess DERA market potential, optimise market participation, design resource portfolio, and establish sustainable business cases.
	
	The proposed framework has been demonstrated through an illustrative canonical example and a real case study based on an actual Australian MV network and market setup. The results highlight the effectiveness of feasibility, ramp, duration and economic NOEs to demonstrate the impact of network, response time, call length and cost on the operation of a DERA, while the capability NOE provides the basic information to measure its overall capacity. In addition, it has been shown that the technical NOE can accurately reflect the technical potential of a DERA to provide a specific service, whereas the commercial NOE could for example be used to construct the DERA bid stack. The comparison between different NOEs also brings unique insights into the role of different stakeholders and relevant use cases in the development of distributed energy marketplaces. Finally, we have also shown the parallelisation and scalability potential of the proposed methodology, which is also useful for computational purposes.
	
	While this paper has focused on fundamental concepts of DER flexibility, which are also being practically used (see e.g.~\cite{ARENA2020}), there is a great deal of ongoing and future work. This includes, for example, investigation and modelling requirements to establish NOEs for unbalanced LV networks in detail. Another important topic is the incorporation of uncertainty in the proposed modelling framework (see e.g.~\cite{Naughton2020} for initial steps), for which potential approaches might include Monte Carlo simulations~\cite{Gonzalez2018}, robust optimisation~\cite{Zhao2016}, multi-parametric programming~\cite{Noto2019}, and so on. 
	\vspace{-1em}
	\section*{Appendix-I}
	\vspace{-0.5em}
	\section*{3D Visualisation of Ramp, Durations and Economic NOEs}
	\vspace{-0.50em}
	For the sake of completeness, the 3D visualisations of the ramp, duration and economic NOEs are presented here.
	
	\vspace{-1em}
	\subsection{Ramp flexibility NOE}
\begin{figure}[]
		\centering
		\includegraphics[width=65mm] {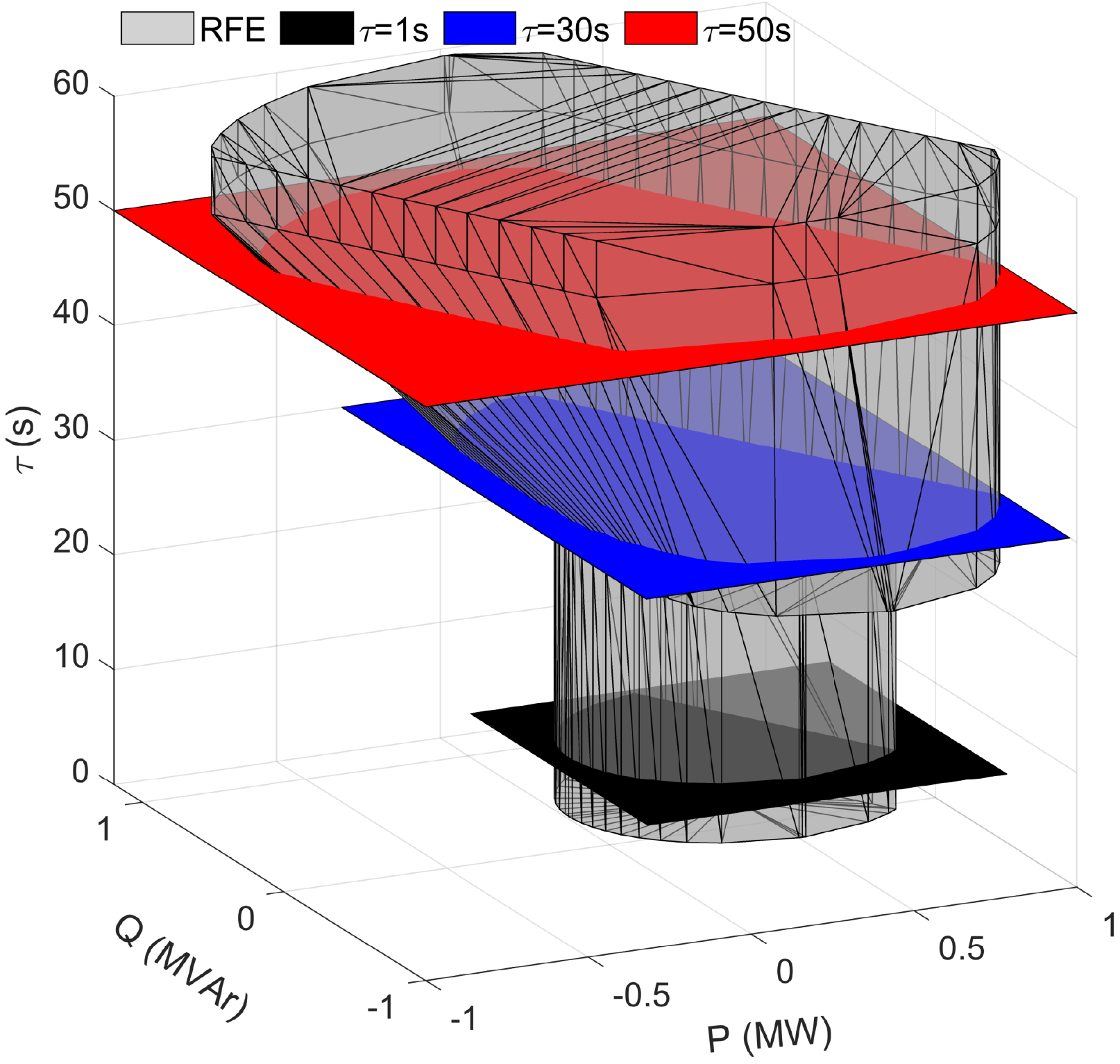}
		\vspace{-1.0em}
		\caption{Ramp flexibility NOE (RFE) visualisation in active power, reactive power and response time ($\tau$) 3D space.}
		\vspace{-01.750em}
		\label{fig:CS_RFE3D}
	\end{figure}

	\vspace{-0.25em}
	The ramp flexibility NOE (RFE) quantifies the DERA speed to alter its operating point and is described through eq.~\eqref{Eq:RampNOE} as a function of active power, reactive power and response time ($\tau$) while considering network constraints. Fig.~\ref{fig:CS_RFE3D} (grey volume) shows the RFE of the test DERA described in Section~\ref{Sec:Exp} and shown in Fig.~\ref{fig:Exp_Net}. Furthermore, the black, blue and red surfaces represent the $\tau$$=$\SI{1}{\second}, $\tau$$=$\SI{30}{\second}, and $\tau$$=$\SI{50}{\second} planes, respectively, and the intersection of RFE and these planes form the iso-ramp contours that are reported in Fig.~\ref{fig:Exp_RFLX} (right).
	Conceptually, these contours could also be calculated for $Q$-$\tau$ and $P$-$\tau$ planes.  
	\vspace{-1em}
	\subsection{Duration flexibility NOE}
\begin{figure}[]
		\centering
		\includegraphics[width=65mm] {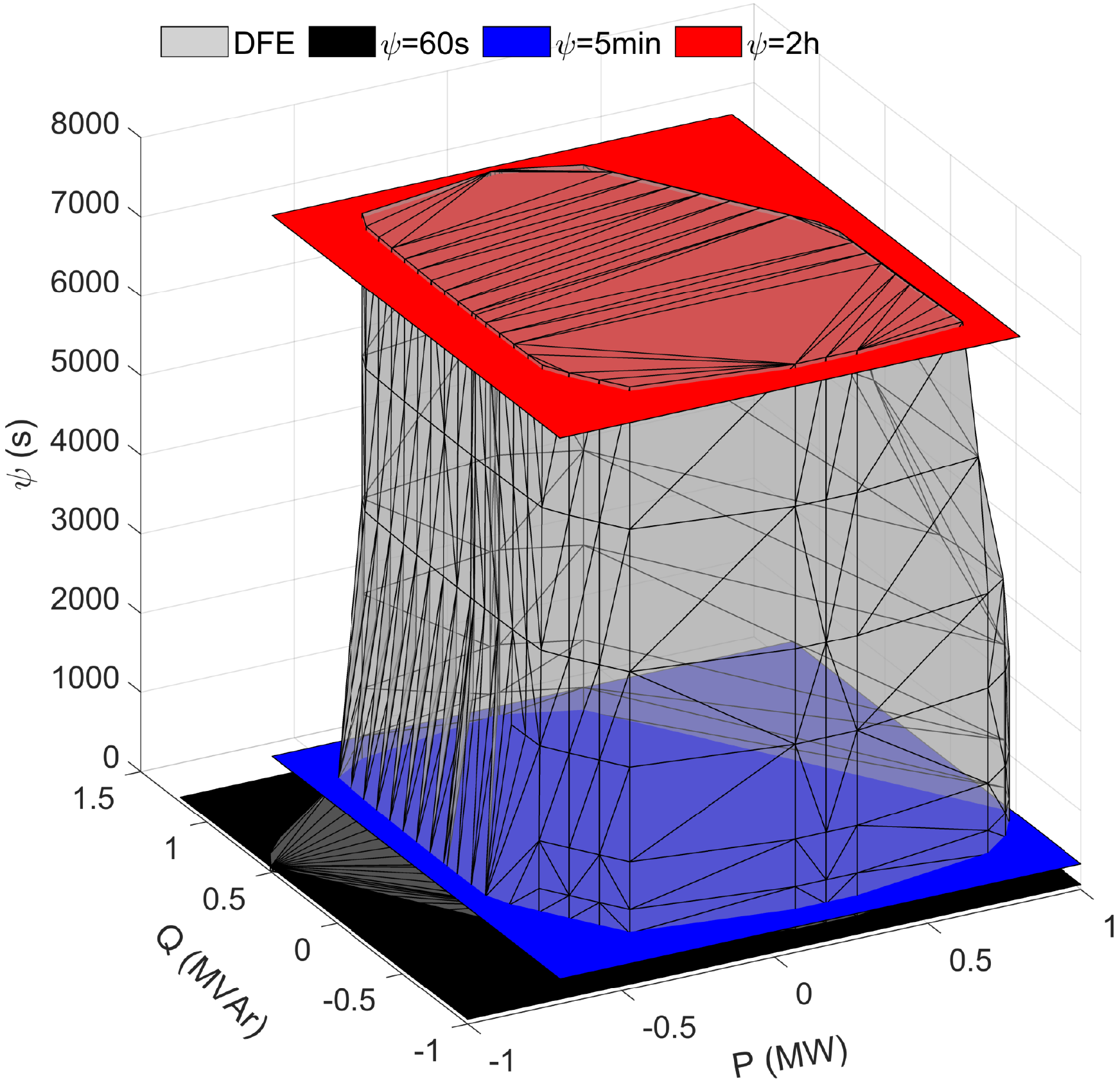}
		\vspace{-1.0em}
		\caption{Duration flexibility NOE (DFE) visualisation in active power, reactive power and service duration ($\psi$) 3D space.}
		\vspace{-0.750em}
		\label{fig:CS_DFE3D}
	\end{figure}

	\vspace{-0.25em}
	The duration flexibility NOE (DFE) quantifies the DERA ability to sustain a response, and is described through eq.~\eqref{Eq:DurationNOE} as a function of active power, reactive power and service duration ($\psi$) while considering network constraints. The DFE of the test DERA described in Section~\ref{Sec:Exp} and shown in Fig.~\ref{fig:Exp_Net} is illustrated in Fig.~\ref{fig:CS_DFE3D} (grey volume), along with the black, blue and red surfaces that represent the $\psi$$=$\SI{6}{\second}, $\psi$$=$\SI{5}{\minute}, and $\psi$$=$\SI{2}{\hour} planes, respectively. The intersection of DFE and these planes form the iso-duration contours that are reported in Fig.~\ref{fig:Exp_DFLX} (right).
	\vspace{-1em}
	\subsection{Economic flexibility NOE}
\begin{figure}[]
		\centering
		\includegraphics[width=65mm] {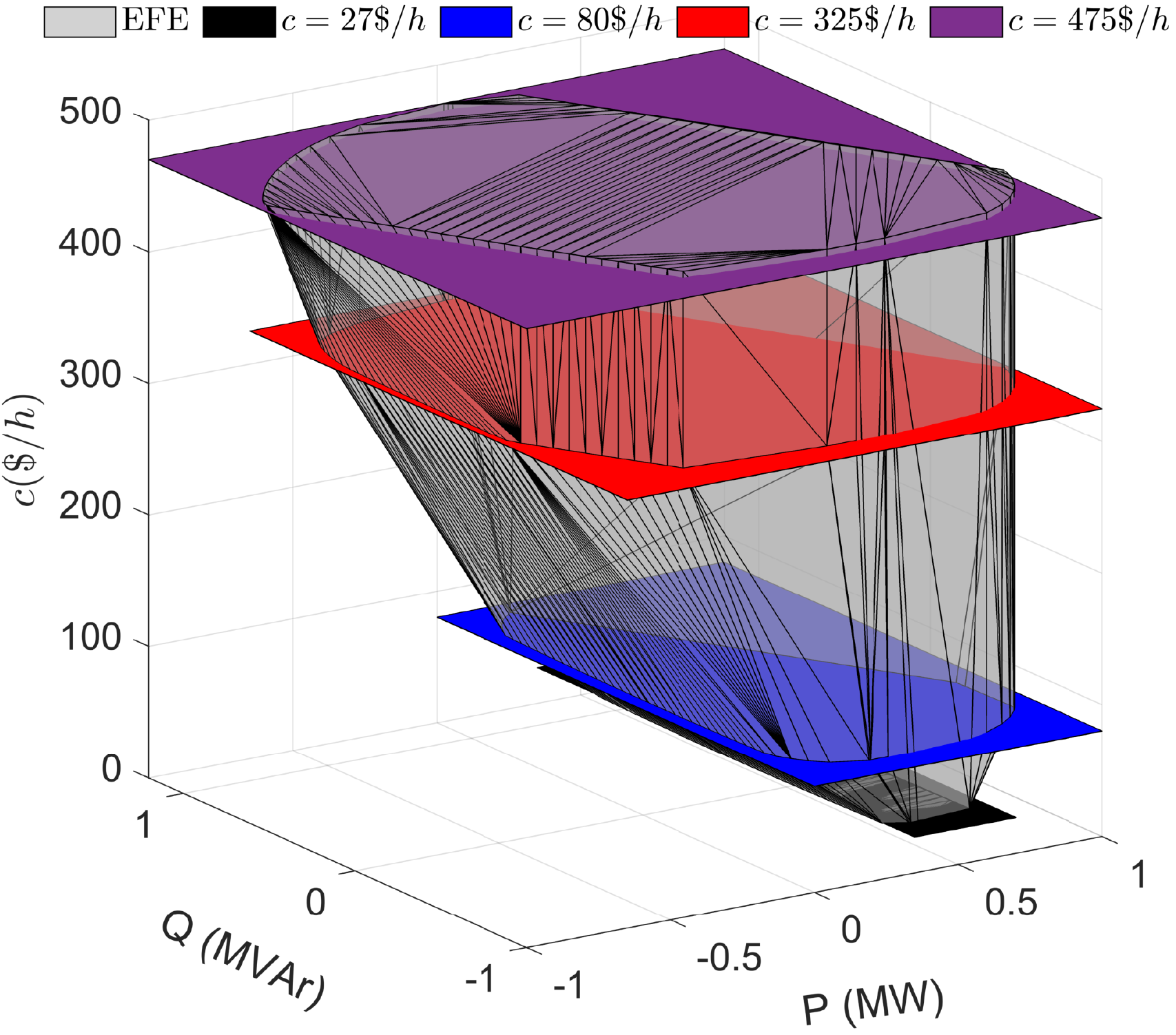}
		\vspace{-1.0em}
		\caption{Economic flexibility NOE (EFE) visualisation in active power, reactive power and cost ($c$) 3D space.}
		\vspace{-1.750em}
		\label{fig:CS_EFE3D}
	\end{figure}

	\vspace{-0.25em}
	The economic flexibility NOE (EFE) reflects the cost incurred to deviate from the DERA operating point, described through eq.~\eqref{Eq:EconomicNOE} as a function of active power, reactive power and cost ($c$) while considering network constraints. The EFE is shown in Fig.~\ref{fig:CS_EFE3D} (grey volume) for the test DERA described in Section~\ref{Sec:Exp} and shown in Fig.~\ref{fig:Exp_Net}. The black, blue, red and purple surfaces represent the $c$$=$\SI{27}{\$\per\hour}, $c$$=$\SI{80}{\$\per\hour}, $c$$=$\SI{325}{\$\per\hour} and $c$$=$\SI{475}{\$\per\hour} planes, respectively. The intersection of EFE and these planes form the iso-cost contours that are reported in Fig.~\ref{fig:Exp_EFLX} (right).
	%\vspace{-0.5em}
%\rev{\section*{Acknowledgements}
%		The authors would like to thank veski for the partial support to conduct this research and John Theunissen from Ausnet Services for the insightful discussions on the topic.}
	\vspace{-1.5em}
	\bibliographystyle{IEEEtran}
	%\vspace{-0.25em}
	%\bibliography{collection}
	\bibliography{mycollection}
\vspace{-1.5em}	
\begin{IEEEbiography}[{\includegraphics[width=1in,height=1.25in,clip,keepaspectratio]{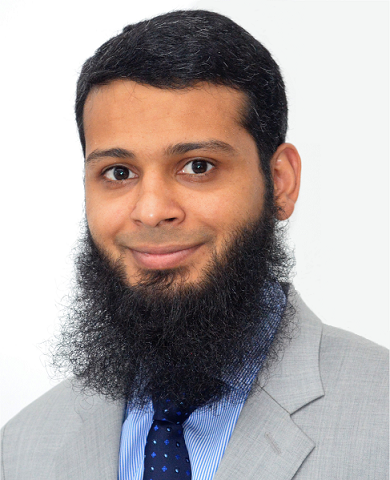}}]%
	{Shariq Riaz} (S'14-M'18-SM'20) received the B.Sc. (Honors) and M.Sc. degrees in electrical engineering from the University of Engineering and Technology Lahore, Pakistan, in 2009 and 2012, respectively, and the Ph.D. degree from The University of Sydney, Australia, in 2018.
	
	He is currently a Research Fellow with the Power and Energy Systems Group at The University of Melbourne, Australia. He has expertise in power system operation and electricity markets. His current research focuses on flexibility from distributed energy resources, multi-energy systems, and distributed energy marketplaces.
\end{IEEEbiography}
\vspace{-1.5em}
\begin{IEEEbiography}[{\includegraphics[width=1in,height=1.25in,clip,keepaspectratio]{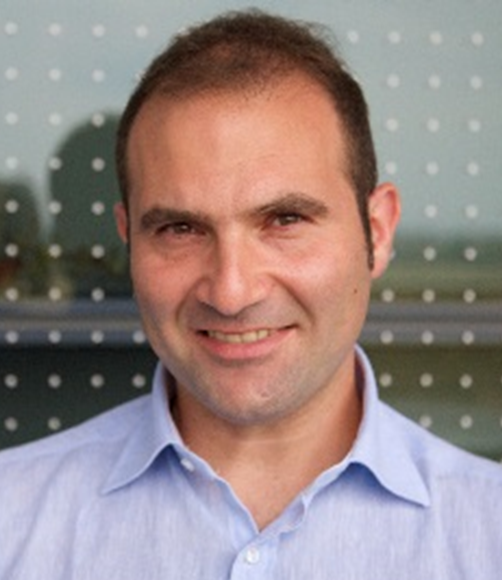}}]%
	{Pierluigi Mancarella} (M’08–SM’14) received the MSc and PhD degrees in electrical energy systems from the Politecnico di Torino, Italy, in 2002 and 2006, respectively.
	
	He is currently Chair Professor of Electrical Power Systems at The University of Melbourne, Australia, and Professor of Smart Energy Systems at The University of Manchester, UK. His research interests include multi-energy systems, grid integration of renewables, energy infrastructure planning under uncertainty, and reliability and resilience assessment of low-carbon networks.
	
	Dr. Mancarella is an Editor of the IEEE Transactions on Power Systems and of the IEEE Transactions on Smart Grid, an Associate Editor of the IEEE Systems Journal, and an IEEE Power and Energy Society Distinguished Lecturer.
\end{IEEEbiography}
	
\end{document}